\documentclass[twocolumn,twocolappendix]{aastex631}
 \usepackage{mathptmx,courier}
  \usepackage[scaled=.92]{helvet}
  \usepackage{CJK}
  \DeclareMathAlphabet{\mathcal}{OMS}{cmsy}{m}{n}
\usepackage{natbib}
\usepackage{graphicx}
\usepackage[figuresright]{rotating}
\usepackage {amsmath,amssymb,eqnarray,tabularx}
\usepackage{enumerate}
\usepackage{enumitem}
\usepackage{mathtools}
\hypersetup{bookmarksnumbered=true,bookmarksopen=true}
\usepackage[all]{hypcap}
\newcommand{\hi}{{\rm H}{\textsc i}}

\def\jyb{\rm{Jy~beam^{-1} }}
\def\jykms{\rm{Jy~km~s^{-1} }}

\def\kms{\rm{km~s^{-1} }}
\def\cmsq{\rm{ cm^{-2} } }

\def\Msun{\rm{M_{\odot}}}
\def\Msunyr{\rm{M_{\odot}} ~{\rm yr}^{-1} }
\def\Msunyrkpcsq{\rm{M_{\odot}} ~{\rm yr}^{-1}~{\rm kpc}^{-2} }
\def\Msunpcsq{\rm{M_{\odot}} ~{\rm pc}^{-2}}
\def\SHI{\Sigma_{\rm HI}}
\def\Ssfr{\Sigma_{\rm SFR}}
\def\Sst{\Sigma_*}

\defcitealias{Wang24}{W24}
\defcitealias{Walter08}{W08}
\defcitealias{Bigiel10}{B10}

\begin{document}

\begin{CJK*}{UTF8}{gbsn}

\title{ FEASTS Combined with Interferometry (II): 
Significantly Changed HI Surface Densities and Even More Inefficient Star Formation in Galaxy Outer Disks}
\correspondingauthor{Jing Wang}
\email{jwang\_astro@pku.edu.cn}

\author[0000-0002-6593-8820]{Jing Wang (王菁)}
\affiliation{ Kavli Institute for Astronomy and Astrophysics, Peking University, Beijing 100871, China}

\author{Xuchen Lin (林旭辰)}
\affiliation{ Kavli Institute for Astronomy and Astrophysics, Peking University, Beijing 100871, China}

\author{Lister Staveley-Smith}
\affiliation{International Centre for Radio Astronomy Research, University of Western Australia, 35 Stirling Highway, Crawley, WA 6009, Australia}
\affiliation{ARC Centre of Excellence for All-Sky Astrophysics in 3 Dimensions (ASTRO 3D), Australia}

\author{Dong Yang (杨冬)}
\affiliation{ Kavli Institute for Astronomy and Astrophysics, Peking University, Beijing 100871, China}

\author{Fabian Walter}
\affiliation{Max-Planck-Institut fu{\" r} Astronomie, K{\" o}nigstuhl 17, D-69117, Heidelberg, Germany}

\author{Zezhong Liang (梁泽众)}
\affiliation{ Kavli Institute for Astronomy and Astrophysics, Peking University, Beijing 100871, China}

\author{Yong Shi (施勇)}
\affiliation{School of Astronomy and Space Science, Nanjing University, Nanjing 210093, China}
\affiliation{Key Laboratory of Modern Astronomy and Astrophysics (Nanjing University), Ministry of Education, Nanjing 210093, China}

\author{Jian Fu(富坚)}
\affiliation{Shanghai Astronomical Observatory, Chinese Academy of Sciences, Shanghai 200030, People's Republic of China}

\author{Hong Guo(郭宏)}
\affiliation{Shanghai Astronomical Observatory, Chinese Academy of Sciences, Shanghai 200030, People's Republic of China}

\author{Luis C. Ho (何子山)}
\affiliation{ Kavli Institute for Astronomy and Astrophysics, Peking University, Beijing 100871, China}

\author{Shu-ichiro Inutsuka}
\affiliation{Department of Physics, Nagoya University, Furo-cho, Chikusa-ku, Nagoya 464-8602, Japan}

\author{Fangzhou Jiang(姜方周)}
\affiliation{ Kavli Institute for Astronomy and Astrophysics, Peking University, Beijing 100871, China}

\author{Peng Jiang(姜鹏)}
\affiliation{National Astronomical Observatories, Chinese Academy of Sciences, 20A Datun Road, Chaoyang District, Beijing, China}

\author{Zhijie Qu(屈稚杰)}
\affiliation{Department of Astronomy and Astrophysics, The University of Chicago, 5640 S. Ellis Avenue, Chicago, IL 60637, USA}

\author{Li Shao(邵立)}
\affiliation{National Astronomical Observatories, Chinese Academy of Sciences, 20A Datun Road, Chaoyang District, Beijing, China}

\begin{abstract}
We update the $\hi$ surface density measurements for a subset of 17 THINGS galaxies by dealing with the short-spacing problem of the original VLA $\hi$ images.
It is the same sample that \citet{Bigiel10} used to study the relation between $\hi$ surface densities and star formation rate surface densities in galaxy outer disks, which are beyond the optical radius $r_{25}$.
For ten galaxies, the update is based on combining original THINGS VLA $\hi$ images with $\hi$ images taken by the single-dish FAST in the FEASTS program. 
The median increment of $\hi$ surface densities in outer disks is 0.15 to 0.4 dex at a given new $\hi$ surface density. 
Several galaxies change significantly in the shape of radial profiles $\hi$ surface densities, and seven galaxies are now more than 1-$\sigma$ below the $\hi$ size-mass relation. 
We update the $\hi$ star formation laws in outer disks. 
The median relation between $\hi$ surface densities and star formation rate surface densities based on pixelwise measurements shifts downward by around 0.15 dex because the $\hi$ surface density values shift rightward, and the scatter increases significantly. 
The scatter of the relation, indicating the star forming efficiency, exhibits a much stronger positive correlation with the stellar mass surface density than before. 
Thus, detecting the previously missed, diffuse $\hi$ due to short-spacing problem of the VLA observation is important in revealing the true condition and variation of star formation possibly regulated by stellar feedbacks in localized environment of outer disks.

\end{abstract}


\section{Introduction} 
\label{sec:introduction}
The relation between neutral gas and star formation is a direct constraint on star formation models, and is used as sub-grid prescription of star formation (SF) process in galaxy formation models \citep{Nobels23}. 
It is also a powerful indicator of the balancing behavior between many physical processes including stellar feedback \citep{Krumholz18, Ostriker22}.
The most direct and important observational demonstration of the relation may be  between the spatially resolved neutral gas surface density and star formation rate (SFR) surface density, typically referred to as the star formation law (SFL) or Kennicutt-Schmidt (KS) law \citep{Kennicutt12}. 
Previous studies found the slope and scatter of the relation to vary significantly with localized condition and among galaxies of different properties.
The slope steepens, while the scatter of SFL increases, toward the low surface densities \citep{Bigiel08}.
The fraction of neutral gas associated with star formation decreases with lower neutral gas surface densities \citep{Roychowdhury11}. 
If separating the neutral gas into different phases, the denser gas tends to exhibit relations with SFR with a slope closer to unity \citep{Bigiel08}. 
The general interpretation is that the the $\hi$ trace the low-column-density gas, serving as the reservoir of star-forming material and feedback energy, while the denser cold molecular gas is the phase closely associated with star formation. 

The galaxy outer disks, roughly beyond the optical radius $r_{25}$, are for many reasons a special location to study star formation.
There, the gravity is weaker compared to the inner disks, leading to a longer dynamic time.
The gas there has lower metallicity and lower surface density, and is dominated by $\hi$ instead of the molecular gas \citep{Leroy08}. 
The gas depletion time is extremely long \citep[\citetalias{Bigiel10} hereafter]{Bigiel10}, H{\textsc i}{\textsc i} regions are much rarer than UV-bright regions \citep{Thilker05}, and the star-forming complexes tend to be small  \citep{Yadav21}.
Moreover, these regions are the expected location for high angular-momentum gas accretion \citep{Grand19}, and are prone to instabilities triggered by high gas fraction and environmental perturbation \citep{Minchev12}. 
$\hi$ warps and optical disk breaks are prevalent in these regions \citep{Pohlen06, vanderKruit07}.
The star formation there is possibly influenced by these dynamic conditions.
Outer disks host star formation with some similarities in the condition and activity to dwarf irregular galaxies \citep{Hunter98, Roychowdhury11} and ultra-diffuse galaxies \citep{KadoFong22}, though their dynamical coldness, dynamic time, and cosmological environment differ systematically.
Characterizing SFL in outer disks is not only for a thorough understanding of star formation, but also important for modeling galaxy evolution, because the outer disks are the frontier of inside-out disk growth and directly link to galaxy structural evolution.

Motivated by observations, modern theories of star formation in low-z galaxies emphasize the multi-phase nature of star-forming gas, and consider the important role of feedback (self)-regulation \citep{FaucherGiguere13}.
The $\hi$ needs to be converted to the molecular gas to feed star formation, and stellar feedbacks work differently on the two phases \citep{Hopkins13}.
In models assuming quasi-equilibrium status of the ISM, the efficiency of converting $\hi$ to the molecular gas depends on the mid-plane pressure with kinematical energy charged by the supernova feedback \citep{Ostriker10}, and$\slash$or on the availability of self-shielding against ultraviolet photons from newly formed massive stars \citep{Krumholz09}. 
Models highlighting the first type of dependence are supported by the observed dynamic pressure equilibrium of the different interstellar medium components with each other and with the gravity \citep{Sun20}.
Models of the latter are supported by the success in reproducing the metal-dependent saturation of $\hi$ column density \citep{Schruba18} as well as the deviation from traditional SFLs of extremely metal-poor dwarf galaxies \citep{Shi18}, because metals strongly absorb ultraviolet photons. 
It is interesting that, while each of these two types of models alone seems to reproduce SFLs in the molecular-dominated inner disks equally well, their physics on feedback regulation become more similar in the outer disks \citep{Krumholz13}. 
It is because in the outer disks, the mid-plane pressure is largely thermal and sustained by ultraviolet photon heating, in contrast to the turbulent inner disks sustained by supernovae \citep{Ostriker11}.

Alternative models are also developed to better reproduce the large scatter in SFL toward low gas densities, possibly applicable to outer disks. 
Non-equilibrium models consider the lag between star formation and feedback \citep{Orr19}, and dynamics-regulated models consider the relative strength of local dynamics and feedbacks \citep{Semenov18}. 
Many of these experimental models focus on single-phase gas, but the simplification may be compatible to the $\hi$-dominated outer disks.
The models discussed above set the SFR by global scale gravity and feedbacks, and are classified as top-down models, in contrast to bottom-up models determining the SFR based on molecular cloud-scale processes \citep{Krumholz14}.
The unique role of $\hi$ as possible turbulence reservoir for the molecular gas is recognized in the bottom-up models, as otherwise the molecular gas collapses too fast.
This is because clumpy molecular gases are embedded in $\hi$, and supersonic motions of cold molecular gases are actually subsonic with respect to the sound-speed of inter-clump medium of $\hi$, and hence turbulent motions of molecular gases avoid shock dissipation \citep{Koyama02, Hennebelle06, Hennebelle19}. 
This effect is potentially more pronounced in the $\hi$-dominated outer disks, where SNe are less frequent and less capable of sustaining the turbulence in the molecular gas. 
The diversity of models is partly due to the limited observational information, particularly the large uncertainty of measurements in the outer disks.

In most of the SF models discussed above, existing stars play a non-negligible role in regulating the SFR, through setting the gravity, metallicity, and the metal-dependent feedback strength. 
The theoretical role of existing stars has been observationally suggested among different galaxy types \citep{Hunter98}, with a fiducial scaling $\Sigma_{\rm SFR}\propto \Sigma_{\rm HI}\Sigma_*^{0.5}$ \citep{Ostriker10, Shi11}. 
The scaling relation including the influence of stellar mass surface densities is referred to as the extended SFL (eSFL for short hereafter) \footnote{ In principle, the KS and extended Schmidt law is the total-neutral-gas SFL. This study focuses on the outer disk where the total neutral gas is dominated by the $\hi$, though traceable molecular gas may still be detected \citep{Schruba11}. Therefore, we refer to the $\Sst$-involved $\hi$-SFL the extended SFL. }.
But it has been also noticed that the slope of star forming efficiency (SFE$=\Ssfr/\SHI$) on kpc-scales steepens in the outer disks in relation not only  to $\hi$ column densities but also to the stellar mass surface densities \citep{Shi18}. 
Such a complex behavior of scaling relation is further complexed by uncertainties in $\hi$ surface density measurements.

One major systematic uncertainty in observing the SFL in outer disks is that the $\hi$ gas, as the major neutral reservoir, is difficult to directly map in traditional interferometric observations. 
This is because the interferometry by nature tends to miss extended fluxes for the lack of sensitivity at close-to-zero baselines. 
The missing of extended structure fluxes most strongly affect the low-surface density outer disks. 
The most promising way to fix the problem is to use total power images taken by single-dish telescopes to supplement the interferometric data \citep{Stanimirovic02, Kurono09, Koda11, Rau19}. 
This is again not easy, as high-resolution single-dish $\hi$ images are rare, and combining the two types of data is more challenging than for CO observations in the millimeter bands, because of the low surface brightness of $\hi$ 21-cm emission lines \citep{Stanimirovic02}. 
Among the many observational efforts on SFLs, the THINGS (The $\hi$ Nearby Galaxy Survey, \citealt{Walter08}, \citetalias{Walter08} hereafter) conducted at the Very Large Array (VLA) has provided a benchmark observational dataset for the kpc-scale $\hi$ measurements.
Unfortunately, these observations also suffer from the aforementioned missing flux problem. 
As part of the FEASTS (FAST Extended Atlas of Selected Targets Survey, \citealt{Wang23}) program, we use the single-dish telescope FAST (Five-hundred-meter Aperture Spherical radio Telescope) to obtain total power $\hi$ images for ten THINGS galaxies within the observable sky of the FAST.
In the past years, there are similar efforts in the literature using the GBT (Green Bank Telescope) for galaxies with large apparent sizes in the sample, which can be more easily resolved by the GBT beam \citep{deBlok14, deBlok18, Eibensteiner23,Lee24}.
The availability of these data suggests that we may derive the $\hi$ surface densities in the outer disks in a more accurate way than before. 

The goal of this paper is thus straightforward. 
We update the $\hi$ column densities for the THINGS sample, which are essential to improving SFLs for calibrating star formation models. 
We also provide new scaling relations in the three dimensional space of $\Sigma_{\rm SFR}$, $\Sigma_{\rm HI}$, and $\Sigma_*$. 
The paper is organized as follows. 
The data used are described in Section~\ref{sec:data}.
The combination of THINGS and FEASTS $\hi$ images, and the deviation of SFR and stellar mass surface density images are described in Section~\ref{sec:analysis}. 
Readers who are mostly interested in the science results can skip this very technical section.
The result are presented in Section~\ref{sec:result}, including the updated $\hi$ surface density distribution (Section~\ref{sec:SHI}), the change in shape of the SFL (Section~\ref{sec:SFL}) and the extended SFL (Section~\ref{sec:exSFL}). 
We discuss the results in Section~\ref{sec:discussion}, and summarize and conclude in Section~\ref{sec:summary}.
The Kroupa initial mass function (IMF, \citealt{Kroupa01}) is assumed throughout the paper. 

\section{Data}
\label{sec:data}
\subsection{Sample}
\label{sec:sample}
The sample of this study are all the 17 spiral galaxies included in \citetalias{Bigiel10}, a benchmark study on the $\hi$ SFL in the galaxy outer disks using $\hi$ images from THINGS. 
The names of these galaxies can be found in the first column of Table~\ref{tab:sample}.
All these galaxies have $\hi$ images taken by THINGS with the VLA (Very Large Array). 
They also have rich multi-wavelength data, of which the most relevant to this study including deep ultraviolet images from GALEX (Galaxy Evolution Explorer) NGS (Nearby Galaxy Survey, \citealt{GildePaz07}), mid-infrared images from Spitzer SINGS (Spitzer Infrared Nearby Galaxies Survey, \citealt{Kennicutt03}) and S$^4$G (The Spitzer Survey of Stellar Structure in Galaxies, \citealt{Sheth10}).

 \begin{table}
    \centering
            \caption{Sample and $\hi$ Measurements}
            \begin{tabular}{c c c c c}
              \hline
              Galaxy  & $\log \, (\frac{M_{\rm HI,W08}}{M_{\odot}})$  &  $R_{\rm HI,W08}$ & $\log \, (\frac{M_{\rm HI,new}}{M_{\odot}})$  & $R_{\rm HI,new}$  \\
              		  & (dex)  & (kpc)  & (dex) & (kpc) \\
              (1)     & (2)     & (3)  & (4) & (5)     \\
                            \hline
NGC 628 & 9.58 & 17.37 & 9.79 & 21.92 \\
NGC 925 & 9.66 & 18.23 & 9.80 & 19.43 \\
NGC 2403 & 9.41 & 14.91 & 9.62 & 17.87 \\
NGC 2841 & 9.93 & 31.33 & 9.99 & 31.08 \\
NGC 2903 & 9.64 & 21.22 & 9.71 & 22.02 \\
NGC 3198 & 9.99 & 32.20 & 10.00 & 32.95 \\
NGC 3351$^a$ & 9.08 & 12.59 & -- & --  \\
NGC 3521 & 9.90 & 23.65 & 9.95 & 24.50 \\
NGC 3621 & 9.84 & 25.35 & 10.06 & 26.70 \\
NGC 3627$^a$ & 8.91 & 7.78 & -- & --  \\
NGC 4736$^a$ & 8.60 & 4.24 & -- & -- \\
NGC 5055 & 9.96 & 20.22 & 10.10 & 21.55 \\
NGC 5194 & 9.40 & 13.97 & 9.59 & 14.56 \\
NGC 5236 & 9.23 & 14.08 & 10.10 & 22.44 \\
NGC 5457 & 10.15 & 32.62 & 10.39 & 39.08 \\
NGC 7331 & 9.96 & 25.95 & 10.04 & 26.49 \\
NGC 7793$^a$ & 10.39 & 7.65 & -- & --  \\

              \hline
            \end{tabular} 
            
              \raggedright
                Column~(1): Galaxy name. Galaxies marked with $a$ do not have new $\hi$ data, and are not updated with new $\hi$ measurements (see Section~\ref{sec:data}). 
                Column~(2): $\hi$ mass from the $I_\text{W08}$ data.
                Column~(3): Characteristic radius for the $\hi$ disk, $R_{\rm HI}$, measured from the $I_\text{W08}$ data. $R_{\rm HI}$ is the semi-major axis of isophotes where $\SHI=1\, \Msunpcsq$. 
                Column~(4): $\hi$ mass measured from the new data.
                Column~(5): $R_{\rm HI}$ measured from the new data.
                {\bf Note:} the superscript $a$ marks the galaxies with relatively small $\hi$ disks for which we do not use new $\hi$ images. 
        \label{tab:sample}
    \end{table}

The $\hi$ images of the sample are updated or unchanged, as listed below
\begin{enumerate}

\item For the two largest galaxies (R$_{\rm HI}>35'$), NGC 2403 and NGC 5236, we take existing $\hi$ images that have been short-spacing corrected using the GBT images \citep{deBlok18, Eibensteiner23} \footnote{ Similar short-spacing correction for NGC 5236 was conducted in \citet{Lee24}.}. 

\item For the next 12 largest galaxies, we correct for short spacing with total-power $\hi$ images from FEASTS for ten of them that are within the observable sky of FAST. 
The THINGS images of these 10 images miss a median of 30\% $\hi$ fluxes \citep[hereafter \citetalias{Wang24}]{Wang24}. 
The two exceptions, NGC 3621 and NGC 7793, are beyond the observable sky of FAST and will be included in the next category. 

\item  For NGC 3621, the natural-weighted $\hi$ image from the LVHIS (Local Volume HI Survey, \citealt{Koribalski18}) is used, which was taken by the ATCA (Australian Telescope Compact Array) less suffering from the short-spacing problem. 
Supportively, its LVHIS $\hi$ flux of 857 $\jykms$ is closer to the single-dish flux of 884 $\jykms$ \citep{Koribalski04} than the THINGS flux of 679 $\jykms$.
This image has relatively large beam major ($b_{\rm bmaj}$) and minor ($b_{\rm min}$) FWHM (full-width-half-maximum), 75$'$ and 43$'$, or 2.4 and 1.4 kpc, respectively. 

\item For NGC 7793 ($R_{\rm HI}=7.4'$), none of the mentioned options are feasible,\footnote{The LVHIS natural-weight image of NGC 7793 has a beam major axis $b_{\rm maj} \sim 6'$.} but its THINGS $\hi$ flux of 246 $\jykms$ is not obviously incomplete in comparison to existing single-dish integrated measurements (271, 197, and 232 $\jykms$ from \citealt{Fisher81}, \citealt{Paturel03} and \citealt{Koribalski04}, respectively). So we use the THINGS-only $\hi$ image for this galaxy. 

\item The remaining three galaxies all have $R_{\rm HI}<3.5'$, much smaller than the critical scale of 8$'$ for THINGS $\hi$ images to significantly miss fluxes \citepalias{Wang24}, so short-spacing is not so crucial for them. We use the THINGS-only $\hi$ images for them. 

\end{enumerate}

In summary, we use {\it new $\hi$ images} for the 13 galaxies that are most likely to suffer from the missing flux problem, among which the data of ten galaxies are newly obtained in FEASTS. 

\subsection{HI Images of FEASTS}
The FEASTS scans $\hi$ in and around nearby galaxies in a uniform observing mode with the FAST. 
The data is reduced with a pipeline developed mostly following standard procedures for single-dish $\hi$ images, but optimized for the FEASTS data (\citealt{Wang23}, W24).
The produced data cubes have a beam FWHM of 3.24$'$, and channel width of 1.65 $\kms$.
They typically have a 3-$\sigma$ column density limit of $5\times10^{17}~ \cmsq$ when assuming a line width of 20 $\kms$. 

The ten galaxies that we will correct for short-spacing with FEASTS data have been studied in W24.
The systematic differences in flux calibration and WCS (World Coordinate System) calibration of these FEASTS images in comparison to the THINGS images have been derived and corrected for each galaxy. 
Moment images for these ten galaxies have also been produced, using masks generated by SoFiA with a threshold of 3.5-$\sigma$ and smoothing kernels that have a maximum size twice the beam FWHM. 

\subsection{HI images of THINGS}
\label{sec:data_things}
The THINGS team obtained $\hi$ images with the VLA, and released the reduced $\hi$ data cubes and moment images \citepalias{Walter08}.
We use the released natural-weighted $\hi$ images (the W08 interferometric dataset). 
They have typical $b_{\rm maj}$ of 10$''$ (0.5 kpc), channel width of 2.6 or 5.2 $\kms$, and a typical 3-$\sigma$ column density limit of $\sim10^{20}~\cmsq$ assuming a line width of 20 $\kms$. 

For the ten galaxies to be corrected for short-spacing with the FEASTS images, we have re-processed with CASA (Common Astronomy Software Applications) the calibrated and continuum-subtracted VLA visibilities to obtain the $\hi$ cubes and images in \citetalias{Wang24}.  
The new processing set a larger field-of-view than the images previously published in \citetalias{Walter08}, to better cover the outskirts of large galaxies.
It used the mutli-scale CLEAN to better account for extended flux, and the natural weighting. 
Like in \citetalias{Walter08}, the CLEAN residual is rescaled to account for difference between dirty and clean beams before added to the clean-beam--convolved model (convolved model for short hereafter), to produce the rescaled cubes. 
The CLEAN residual is also directly added to the convolved model to produce the standard cube, in which the fluctuation of pixel values in the blank region reflect the rms level.
Source masks have been obtained from standard cubes using SoFiA with similar smoothing and thresholding settings as in \citetalias{Walter08}, and moment images and fluxes are derived using these masks and the rescaled cubes.
These newly processed images have consistent beam size, depth, integral fluxes and amplitude spectral shape in the Fourier space (when the same field-of-view is selected) in comparison to the \citetalias{Walter08} images, as shown in \citetalias{Wang24}. 

The newly processed THINGS images for the ten galaxies from the FEASTS subset, plus the W08 images for the rest 7 galaxies, consist the {\it W24 interferometric dataset}. 
All THINGS images are corrected for the VLA primary beam attenuation. 

\subsection{GALEX Ultraviolet Images and Spitzer Mid-infrared Images}
The GALEX NGS FUV images typically have a point spread function (PSF) FWHM of 4.5$'$, and background rms of $\sigma \sim$28.86 mag arcsec$^{-2}$ in surface brightness. 
The Spitzer IRAC 1 (3.6 $\textmu$m) images are taken from the SINGS when available, and otherwise from the S$^4$G for two galaxies, NGC 2903 and NGC 5457.
These two types of images have a PSF FWHM of 1.67$''$ and 2.1$''$, and typical background rms of 25.7 and 25.6 mag arcsec$^{-2}$,  respectively.

\section{Analyses}
\label{sec:analysis}

\subsection{Combining the THINGS and FEASTS HI images} 
\label{sec:comb}
\begin{figure*} 
\centering
\includegraphics[width=16cm]{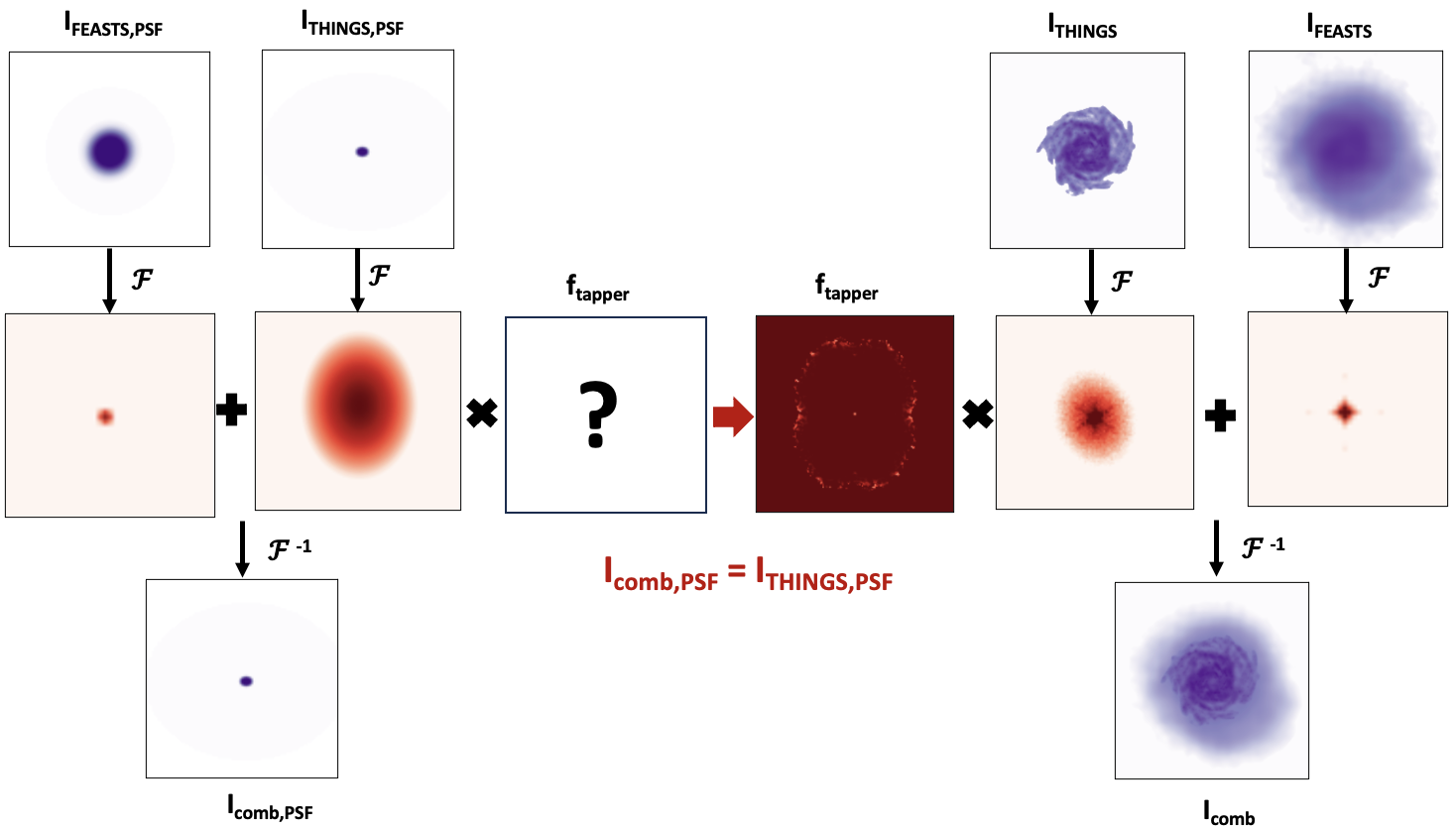}
\caption{ Work flow of the image combination procedure. The part illustrated in this figure starts with THINGS (I$_{\rm THINGS}$) and FEASTS (I$_{\rm FEASTS}$) images that are projected to the same WCS coordinates, and multiplied with the VLA primary beam.  The $\mathcal{F}$ and $\mathcal{F}^{-1}$ marks mean the Fourier and inverse Fourier transformation, respectively. The 2D weighting function $f_{\textrm{tapper}}$ is solved by requiring the linearly combined PSF to be equivalent to the THINGS PSF (left side). It is then applied to the procedure on the right side to produce the combined images (I$_{\rm comb}$).  Motivated by a similar figure in \citet{Stanimirovic02}. }
\label{fig:workflow}
\end{figure*}

We combine the THINGS moment-0 images with the FEASTS ones, using the latter to correct for the short-spacing of the former. 
We adopt the linear combination method \citep{Stanimirovic02}, and develop a python code that has similar function as the CASA script {\it feather} and Miriad script {\it immerge}. 
The procedure is explained below, and the key steps are illustrated in Figure~\ref{fig:workflow}.
For each galaxy, the procedure firstly project the FEASTS image to the WCS coordinates of the THINGS image. 
The pixel value units of both THINGS (primary beam corrected) and registered FEASTS images are converted from $\jyb$ to Jy pix$^{-1}$.
These two images are multiplied with the VLA primary beam to recover the image domain response of the VLA observation. 
They are then transformed to the Fourier space, which produces the FFT images.
The THINGS FFT images are multiplied with a weighting function $f_{\textrm{tapper}}$ before being added to the FEASTS FFT image.
The weighting function $f_{\textrm{tapper}}$ follows that adopted by {\it immerge}, and is calculated as $1-f_{\rm FEASTS}/ f_{\rm THINGS}$, where $f_{\rm FEASTS}$ and $f_{\rm THINGS}$ are the peak flux normalized (to unity) beam (PSF) images of the FEASTS and THINGS data in the Fourier space, respectively.
This $f_{\textrm{tapper}}$ helps to obtain a final PSF of the combined image that is close to the THINGS PSF. 
The added FFT image is inverse Fourier transformed, multiplied with the THINGS beam area to recover the unit of $\jyb$, and divided by the VLA primary beam, to produced the final, combined image. 
By construction, each combined $\hi$ image has the same spatial resolution as the corresponding THINGS one, and the same total flux as the corresponding FEASTS one.

In \citetalias{Wang24}, the moment-0 images of the two types of data have been compared in the Fourier space in overlapping spatial frequencies to cross-calibrate the fluxes to the THINGS level.
This step is necessary before revealing any excess $\hi$ detected in FEASTS data, and linearly combining the two types of data.
The moment-0 images instead of the cubes are used for the combination.
It is a natural extension of the flux-level cross-calibration procedure based on the moment-0 images.  
Mathematically, it is equivalent to a longer procedure of firstly using the source mask of each data cube to blank noise dominated regions, then combining the blanked data cubes, and finally making moment images from the combined cube.
Using separate masks for the THINGS and FEASTS data cubes has the risk of missing fluxes beyond masks in different ways, which should be responsible for the unsmooth transition (between THINGS-detected and FEASTS-only regions) in the combined moment-0 images (Figure~\ref{fig:atlas}). 
However, without the necessary source masks, particularly those for the THINGS cubes, low-density fluxes revealed by FAST would be buried in the higher noise of THINGS data.
For the THINGS moment-0 images, we use those from the convolved model cubes produced in the CLEAN procedure in \citetalias{Wang24} (see section~\ref{sec:data_things}), instead of from the standard or rescaled cubes. 
We justify this choice in Appendix~\ref{app:comb}, which is mainly based on the robustness of pixelwise measurements of low $\hi$ surface densities ($\sim$1 $\Msun$) in mock data.
These settings produce the {\it W24 combined dataset}. 
The $\hi$ column density images derived from the combined moment-0 images, in comparison to the THINGS and FEASTS column density images, are displayed in Figure~\ref{fig:atlas}.

For comparison of analysis, we also combine the W08 moment-0 images with the FEASTS moment-0 images, to produce the {\it W08 combined dataset}.  
We note that the W08 moment-0 images are generated from rescaled cubes. 
With mock tests, we find that using moment-0 images from rescaled cubes in image combination should produce as good radial profiles of $\SHI$ as using moment-0 images from convolved models (Appendix~\ref{app:comb}). 
Radial profiles of $\SHI$ measured from the W08 data combined images will be compared to those from the W24 combined images in Section~\ref{sec:profile}.

\subsection{SFR and Stellar Mass Surface Density Maps}
The GALEX NGS FUV images are used to derive the SFR surface densities ($\Ssfr$). 
FUV emission mostly from O and B stars can be used to trace unobscured SFRs with a typical timescale of $\sim20\,\text{Myr}$ \citep{Salim07}.
Photometry of the GALEX images are conducted using a pipeline with details described in \citet{Wang17}, but have been translated in to python scripts, and incorporated the latest improvements of \emph{astropy} \emph{photutils}. 
The source masks and background images of GALEX FUV images are produced. 
The background is subtracted, and the sources other than the target galaxy is blanked in each image.
The foreground stars on top of target galaxies are identified and masked, based on the $\text{FUV} - \text{NUV}$ color, following the criteria of \citet{Leroy08}.
We correct for Galactic extinction using the SDF dust map \citep{Schlafly11} available at the IRSA Dust Extinction Service, assuming A$_{\rm FUV}=$ 8.24 E(B$-$V) \citep{Wyder07}. 
The SFR in each pixel is estimated from the background-subtracted FUV images using the equation 2 of \citetalias{Bigiel10}, which is an IMF converted version of the equation 10 in \citep{Salim07}. 
Following B10, we do not correct for the internal dust attenuation unless specified, but caution its possible dependence on the $\hi$ column density. 

The Spitzer IRAC 1 ($3.6\,\textmu$m) images are used to derive the stellar mass surface densities ($\Sst$).  
The $3.6\,\textmu$m emission is largely from photospheres of evolved stars in star forming galaxies, but contaminations from polycyclic aromatic hydrocarbon (PAH), very hot dusts, and intermediate-age stars can sometimes be considerable \citep{Querejeta15}.
Photometry of the Spitzer images is conducted using the pipeline described in Liang et al. (in prep).
 The source masks, background maps, and uncertainty maps are produced for the SINGS images, and directly taken from the release of \citet{MunozMateos13} for S$^4$G images.
 We remove the background and sources other than the target galaxy. 
The Galactic extinction is corrected assuming A$_{3.6\,\text{\textmu{}m}} =$0.17 E(B$-$V) \citep{Jarrett13}. 
The stellar mass in each pixel is firstly estimated from the background-subtracted IRAC 1 images using the equation 5 of \citet{MunozMateos13}, and then divided by 1.6 to convert the IMF from \citet{Salpeter55} to \citet{Kroupa01}. 
Through separating the emission of old stars from that of contaminants, \citep{Querejeta15} found good agreement with this equation on the $3.6\,\textmu$m stellar mass-to-light ratio.

The maps for $\Ssfr$ and $\Sst$ are further obtained by dividing the pixelwise quantities with the pixel area, and multiplying with the optical axis ratio \citepalias{Walter08} to correct for projection.
The typical pixelwise 1-$\sigma$ $\Ssfr$ depths range from 10$^{-5.2}$ to 10$^{-4.2}$ $\Msunyr$ throughout the sample, and is $\sim10^{-4.7} \Msunyrkpcsq$ on the median. 
The pixelwise 1-$\sigma$ $\Sst$ depths range from 0.3 to 11 $\Msunpcsq$ throughout the sample, and $\sim$1 $\Msunpcsq$ on the median.

We notice that our measurements of the $\Ssfr$ are systematically higher than those in B10 (see Section~\ref{sec:SFL}), particularly for large galaxies including NGC 5194. 
The difference is possibly due to the difference in details in the FUV photometry. 
For example, we notice that B10 estimated the image background as the median of residual pixel values, after discarding emission with intensities greater than 3-$\sigma$ above the median value of the image.
We instead generate a detection mask after smoothing each image with a 2D gaussian kernel that is with an FWHM eight times of PSF and setting a 2-$\sigma$ detection threshold. 
We estimate the background as the 3-$\sigma$ clipped mean of the residual pixels outside the detection mask. 
Our treatment can be more efficiently in excluding the faint and extended FUV emissions in the outer disks of large galaxies.
To avoid the interference of systematic differences from simultaneously changing both the $\SHI$ and $\Ssfr$, throughout this work, we keep using the $\Ssfr$ value from our (new) measurements.
That is, when comparing the SFLs, only the $\SHI$ varies between being measured from W08 and updated images.

In the past decades, methods that empirically combine multiple bands or utilize spectral energy distribution (SED) fitting have become very useful in estimating the SFR and stellar mass, by better constraining the influence of dust attenuation, star forming history, and illuminating source \citep{Conroy13}. 
We have instead used a relatively simple method, linearly converting single-band luminosities to these properties, partly to be in line with the SFR estimation used in B10 as our primary purpose is to update the SFL with more accurate $\hi$ surface densities.
Additionally, in outer disks, the SNRs in most bands are low, and the systematic influence of observational and physical factors (e.g. the confusion noise in Spitzer bands, the heating of old stars on the dust) are not well constrained \citep{Eskew12, Leroy12}.
Following the argument in \citetalias{Bigiel10}, it is more straightforward and less model dependent to keep the property estimation closer to observables. 

%

\subsection{Registering and Smoothing Images}
We project the $\Sst$ and $\Ssfr$ images to the WCS system of the combined $\hi$ images, and smooth all images to the same resolution of PSF FWHM$=1$ kpc, so that surface densities of different properties are measured coherently from the same regions. 
The smoothing is done by convolving each image with a Gaussian kernel, for which the FWHMs along the major and minor axis are the square root of difference in quadrature between the targeted PSF FWHM and the original PSF FWHM of the image.
For one galaxy, NGC 3621, the new $\hi$ image has $b_{\rm maj}$ and $b_{\rm min}$ (2.4 and 1.4 kpc, respectively) exceeding 1 kpc, and we skip the smoothing step for this $\hi$ image.

We also conduct a separate set of analysis by smoothing images to a common resolution of 15 arcsec, following the treatment in B10. 
Two galaxies among the FEASTS subset, NGC 2903 and NGC 3521 have the PSF FWHM of their $\hi$ images larger than 15-arcsec (by $\lesssim$10\%), so these $\hi$ images are left un-smoothed. 
As for the new $\hi$ images for NGC 2403 and NGC 5236, which was short-spacing corrected in the literature, we skip the smoothing step for the same reason. 
The 15-arcsec version of results are only presented for direct comparison with B10 in Section~\ref{sec:SFL}. 

\subsection{Surface Density Measurements in Outer disks}
\label{sec:Sigma_measure}
We focus on measurements in the outer disks (beyond $r_{25}$, as in B10), where $r_{25}$ is the major axis of the 25 mag arcsec$^{-2}$ isophote in the B-band taken from \citepalias{Walter08}. 
The surface densities are measured in two ways, (1) as averaged values in pixels with sizes equivalent to the resolution FWHM, and (2) as azimuthally averaged values in elliptical annuli. 
The former way better exhibits scatters in relations due to the diverse conditions and stages of star formation, while the latter is better for capturing the major trend, by averaging out the spatially and temporally stochastic behavior. 
The averaging pixels for pixelwise measurements are taken from a wide annulus with $r_{25}$ and 2 $r_{25}$ as the inner and outer semi-major axis, following the selection of B10.
The azimuthally averaged measurements for SFLs are along a series of narrow annuli, the major axes of which range from 0.5 to 2 $r_{25}$, with an increasing step of 0.1 $r_{25}$.
For $\hi$ radial profiles presented in Section~\ref{sec:profile}, we further measure azimuthally averaged $\SHI$ throughout the $\hi$ disks. 
All elliptical annuli mentioned above have the same center, position angle, and axis ratio as the optical disk \citepalias{Walter08}.
Undetected $\Ssfr$ pixels are assigned $\Ssfr$ values of $10^{-6}\, \Msunyrkpcsq$ in the pixelwise measurement dataset, which is far below the 1-$\sigma$ threshold of $10^{-4.7}\, \Msunyrkpcsq$ for individual pixels among the sample. 
When obtaining the results in Section~\ref{sec:result}, we select statistical methods (e.g. deriving percentiles of $\Ssfr$ in $\SHI$ bins) to minimize selection effects due to these undetected pixels.


\section{Results}
\label{sec:result}
\begin{figure*} 
\centering
\includegraphics[width=16cm]{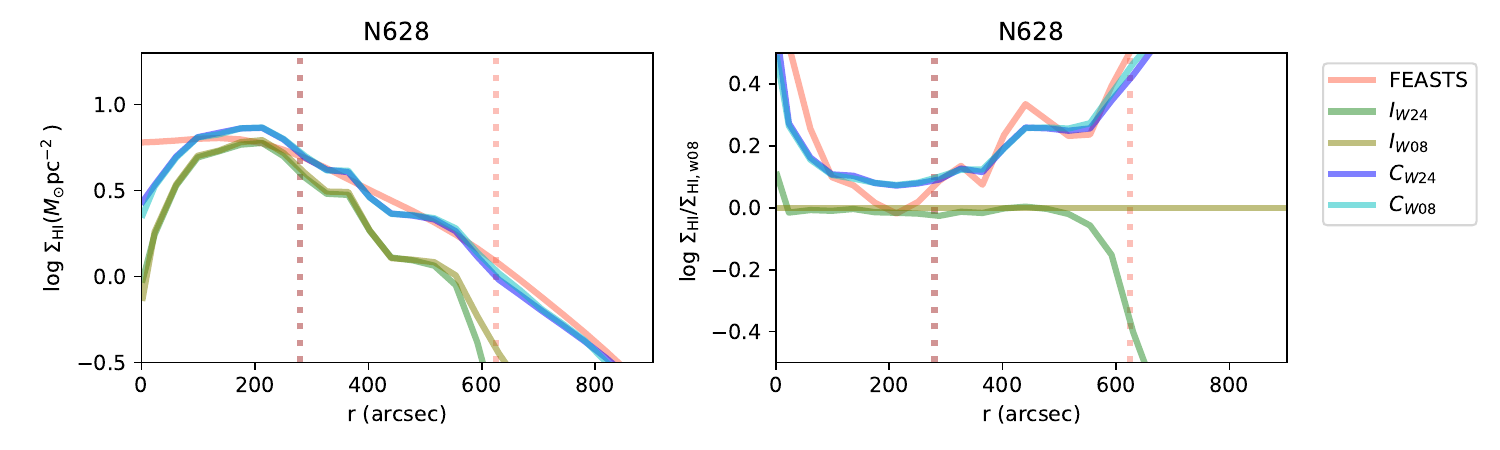}
\includegraphics[width=16cm]{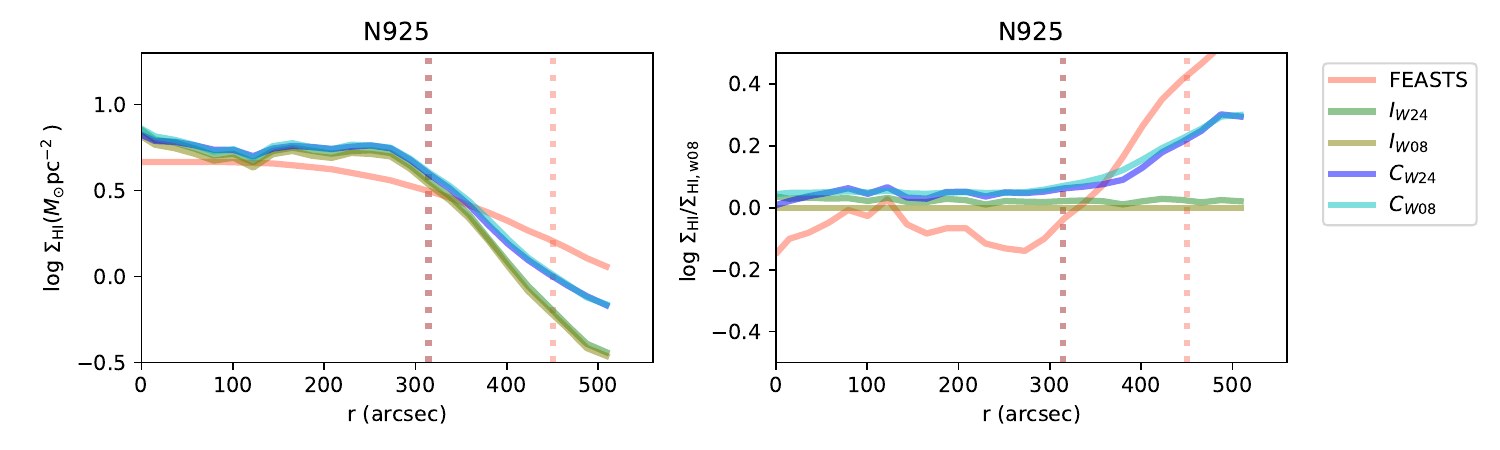}
\includegraphics[width=16cm]{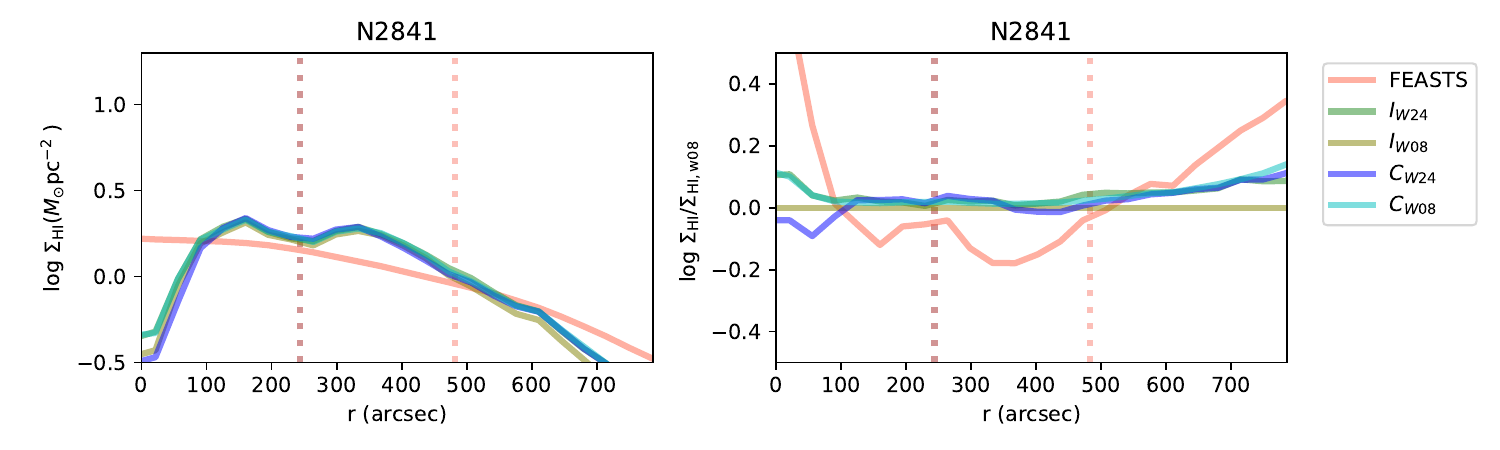}
\includegraphics[width=16cm]{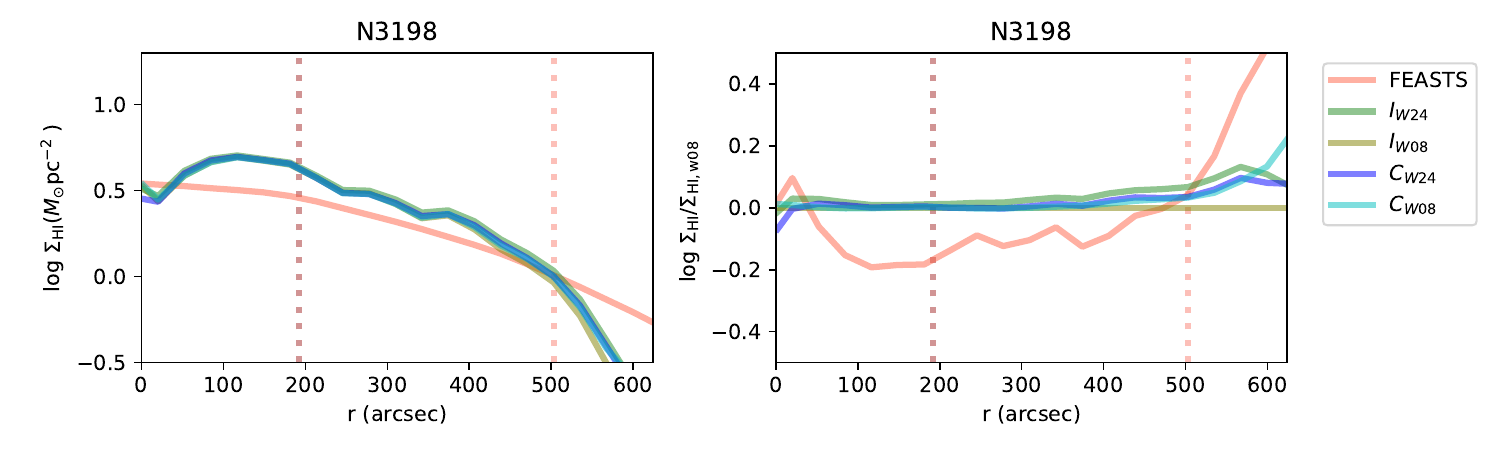}

\caption{ Comparing radial profiles of $\SHI$ measured from different images for galaxies in the FEASTS subset. 
Each row is for a galaxy, with the left panel showing the radial profiles and the right panel the profile difference with respect to the $I_\text{W08}$ measurements. 
For each galaxy, the comparison involves two sets of THINGS-only measurements, $I_\text{W08}$ (olive) and $I_\text{W24}$ (green), based on the W08 and W24 residual-rescaled data cubes respectively.  
It also involves two sets of combined images, $C_\text{W08}$ (cyan) and $C_\text{W24}$ (blue), using the W08 residual-rescaled image and the W24 convolved model to combined with the FEASTS image respectively.  
It also includes profiles directly measured from the FEASTS images (red).
The pink and brown dotted vertical lines mark the positions of $R_{\rm HI}$ (derived from $C_\text{W24}$ profile) and $r_{25}$ respectively. 
To be continued.
}
\label{fig:prof1}
\end{figure*}
\addtocounter{figure}{-1}
\begin{figure*} 
\centering
\includegraphics[width=16cm]{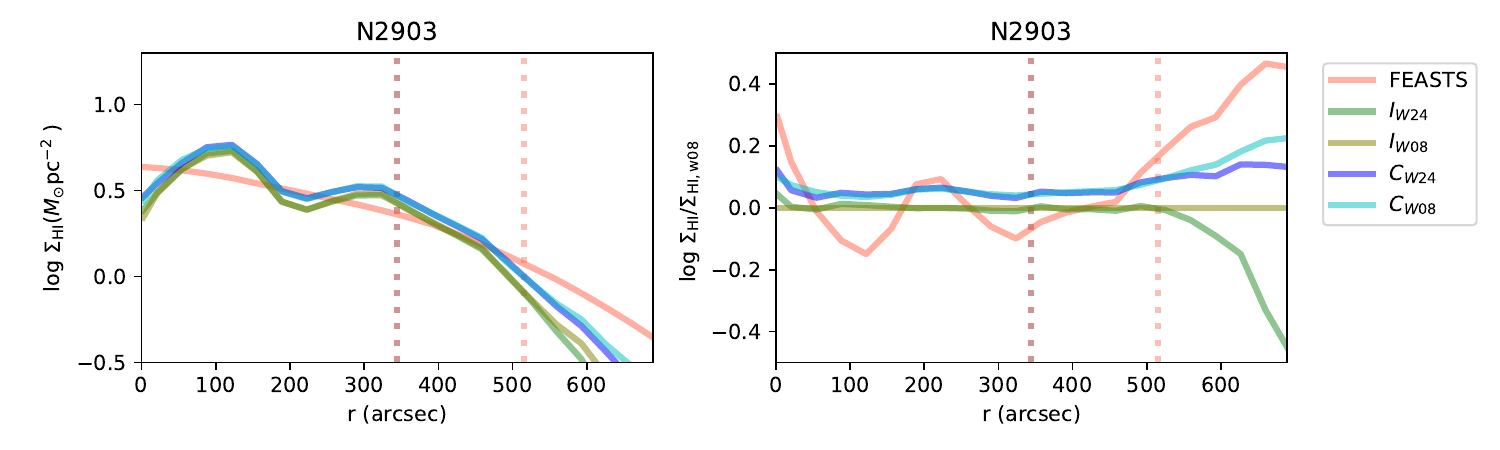}
\includegraphics[width=16cm]{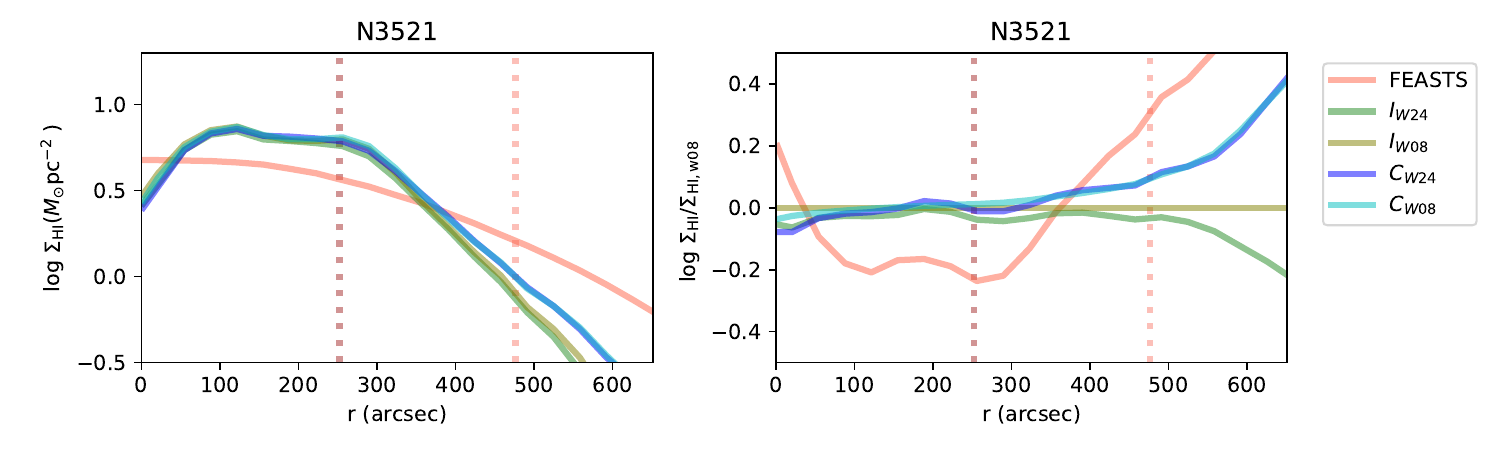}
\includegraphics[width=16cm]{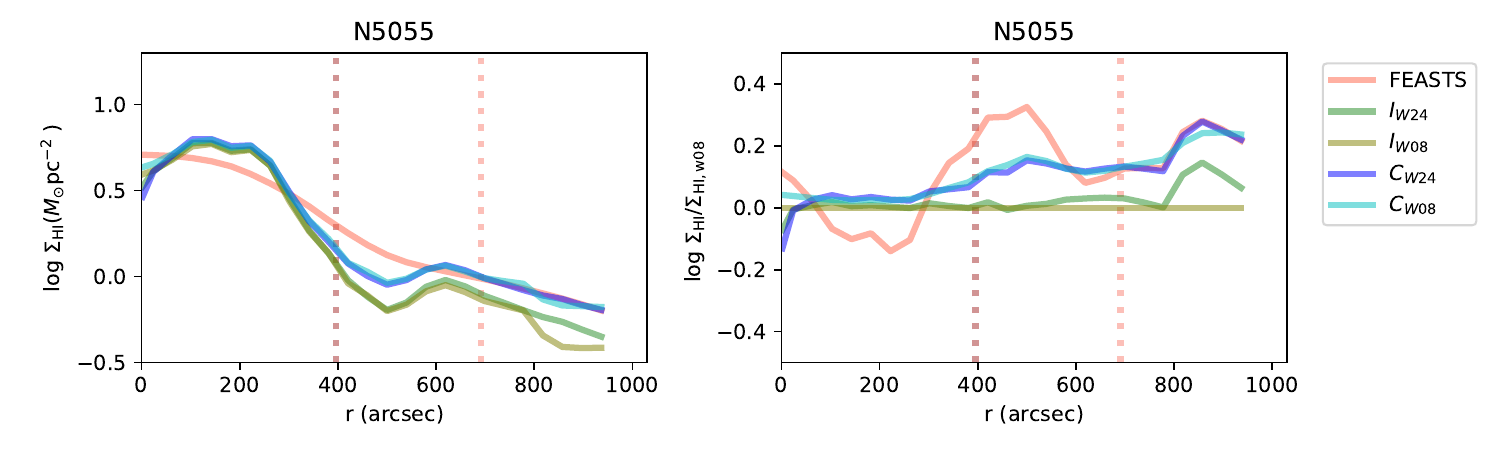}
\includegraphics[width=16cm]{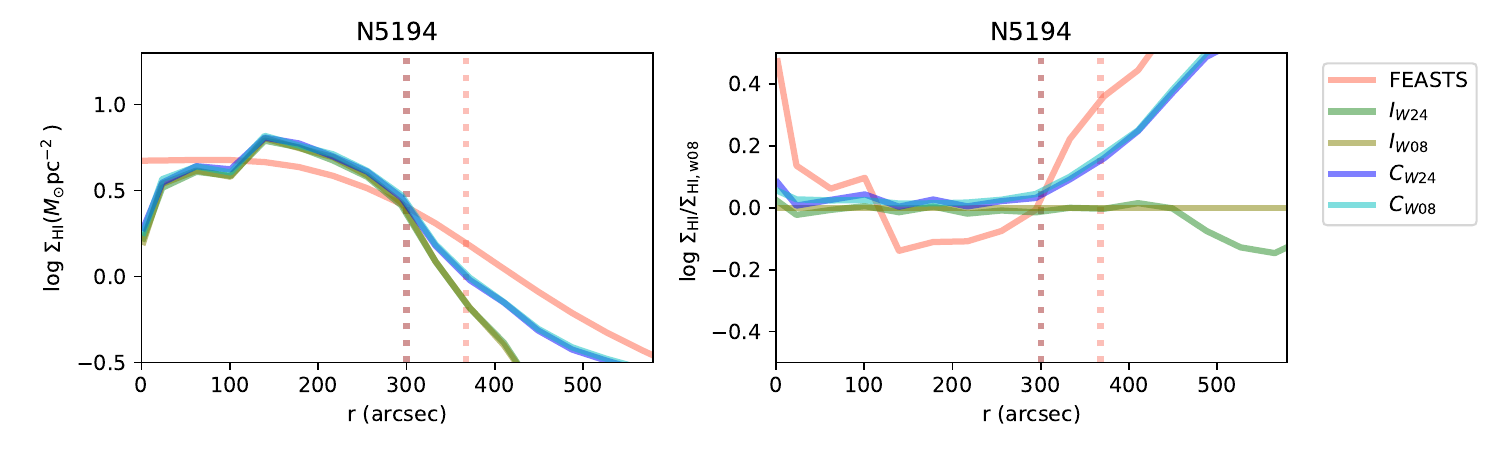}
\caption{Comparing radial profiles of $\SHI$ measured from different images for galaxies in the FEASTS subset. 
Continued.  }
\label{fig:prof1}
\end{figure*}

\addtocounter{figure}{-1}
\begin{figure*} 
\centering
\includegraphics[width=16cm]{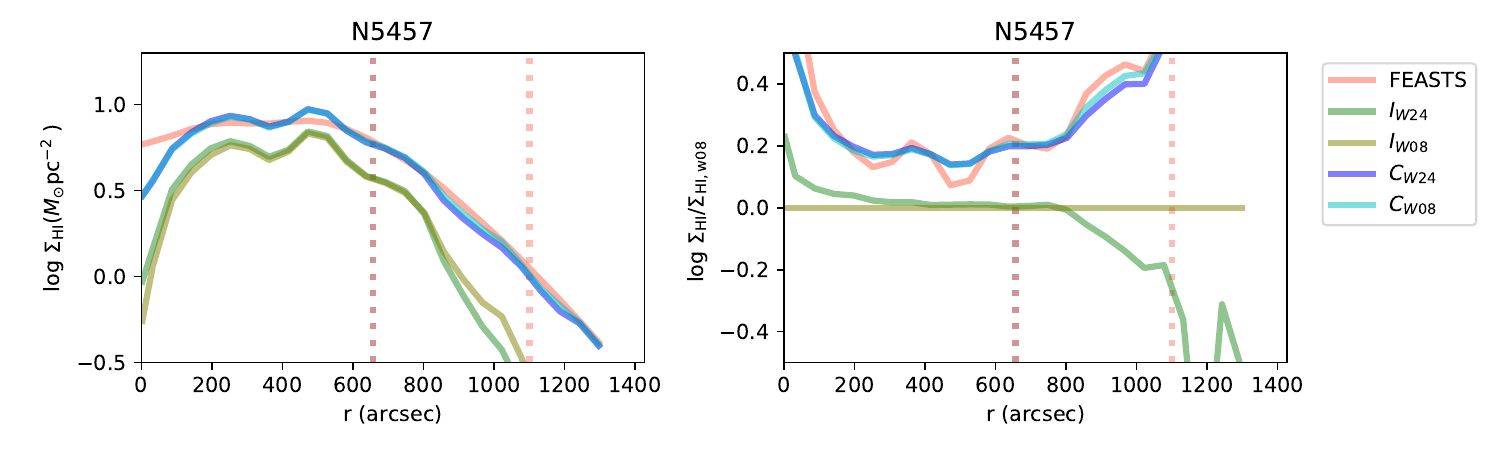}
\includegraphics[width=16cm]{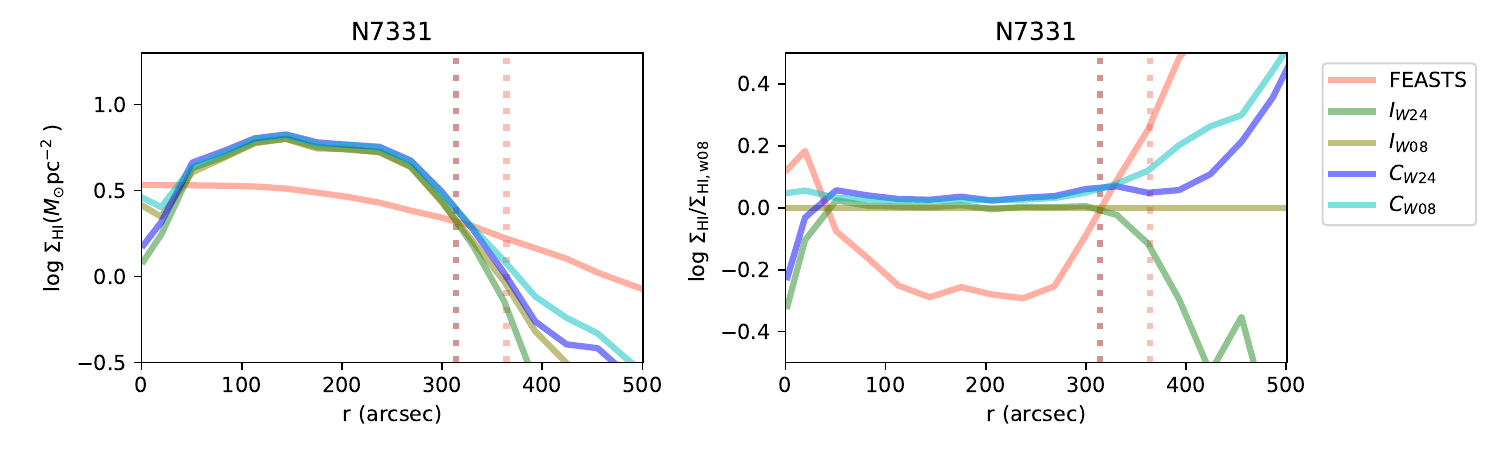}
\caption{Comparing radial profiles of $\SHI$ measured from different images for galaxies in the FEASTS subset. 
Continued.  }
\label{fig:prof1}
\end{figure*}

In this section, we present $\hi$-surface-density-related results, focusing on a comparison of results from the updated and W08 THINGS $\hi$ images. 
Results from the W24 combined images (i.e. W24 convolved model$+$FEASTS) are labeled as $C_\text{W24}$, while those from the W08 combined images (W08 rescaled images$+$FEASTS) are labeled as $C_\text{W08}$. 
Results from the W08 images and W24 residual-rescaled interferometric images are labeled as $I_\text{W08}$ and $I_\text{W24}$, respectively. 
Results labeled by ``Updated'' or without specific tags are based on the updated dataset (the combination of $C_\text{W24}$, images from the literature including \citealt{deBlok18, Koribalski18} and \citealt{Eibensteiner23}, and $I_\text{W24}$ for the 4 smallest galaxies, see section~\ref{sec:data}).


\subsection{Updated HI Surface Densities}
\label{sec:SHI}
\subsubsection{The $\SHI$ Profiles}
\label{sec:profile}

\begin{figure*} 
\centering
\includegraphics[width=16cm]{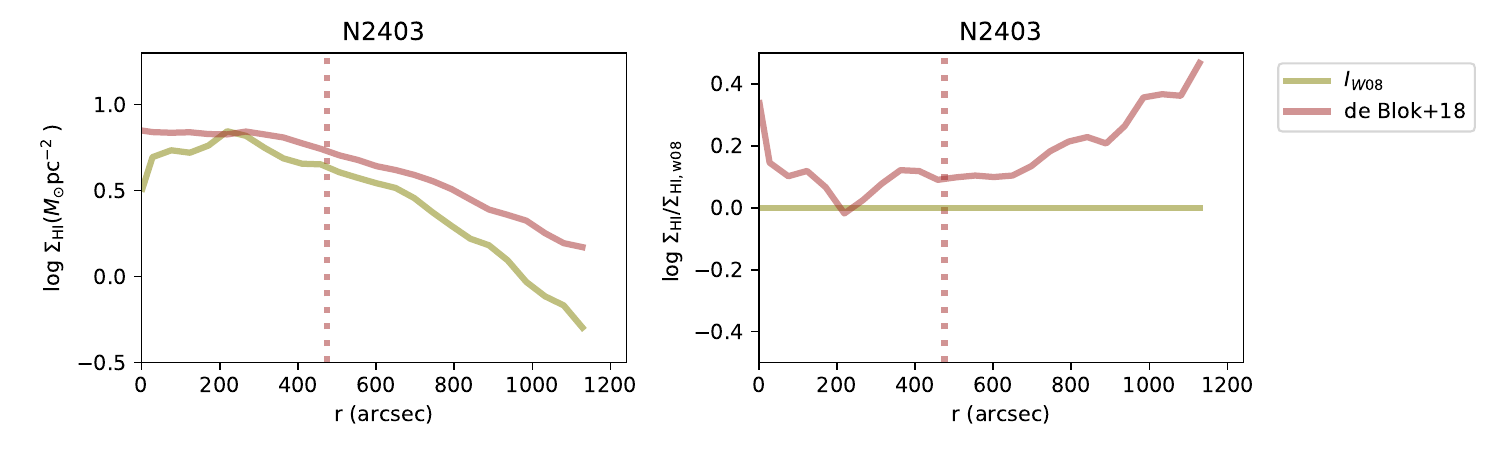}
\includegraphics[width=16cm]{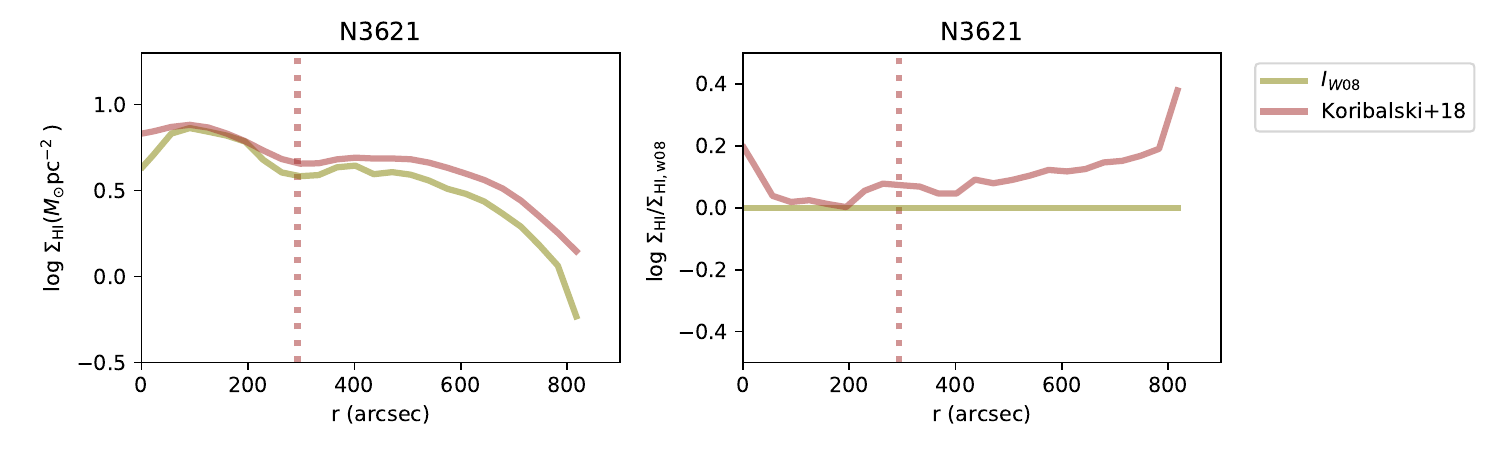}
\includegraphics[width=16cm]{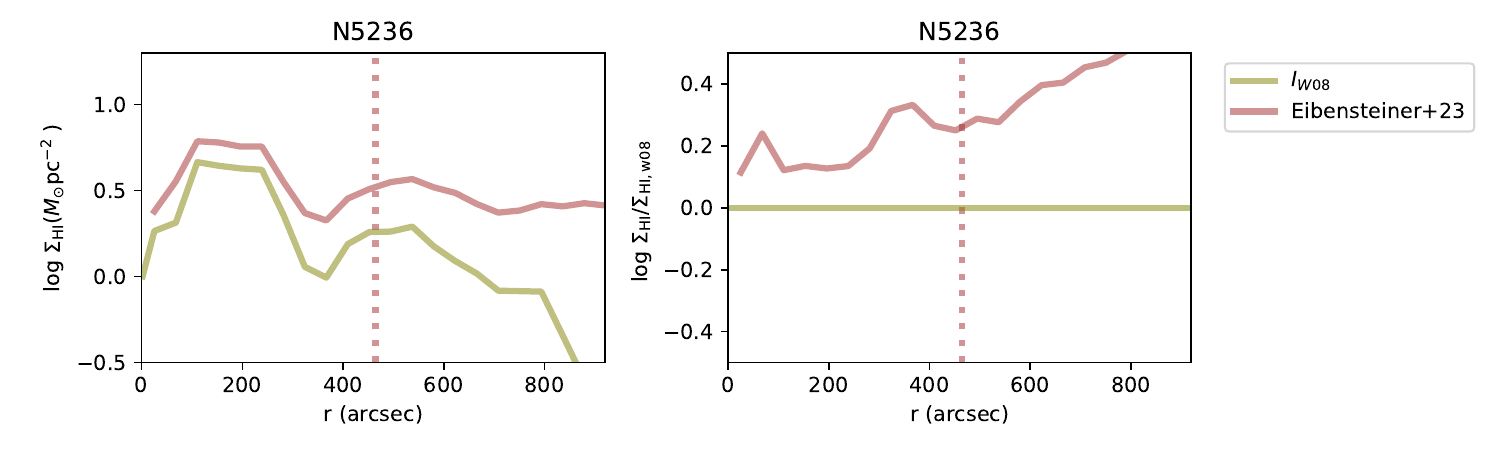}
\caption{ Comparing radial profiles of $\SHI$ measured from different images for galaxies with new $\hi$ images in the literature but not in the FEASTS subset.  
Plotted in a similar way as in Figure~\ref{fig:prof1}, but the $I_\text{W08}$ profiles are only compared with profiles measured from new images from reference papers including \citealt{deBlok18, Koribalski18} and \citealt{Eibensteiner23} (brown) as labeled in the plots (also see Section~\ref{sec:data}.) }
\label{fig:prof2}
\end{figure*}

We firstly directly compare the $\hi$ surface densities of the updated $\hi$ images with those of the THINGS images. 
Figure~\ref{fig:prof1} compares the radial profile shapes for the subset of galaxies having FEASTS data. 
The left panels directly show the profiles of different types, while the right panels more clearly demonstrate the differences with respect to the $I_\text{W08}$ profiles.

The $I_\text{W24}$ and $I_\text{W08}$ profiles are on the whole similar, particularly in the inner disks, supporting our mixed use of the two types of the data. 
The only exception is NGC 7331, which has a 0.2 dex difference in the central $\SHI$ between the two types of profiles. 
The differences in the outer disks of many galaxies and central region of NGC 7331 indicate systematical uncertainties in the interferometric $\SHI$ due to different CLEAN details.

The $C_\text{W24}$ and $C_\text{W08}$ profiles trace each other closely throughout the disks.
Their similarity holds even at the presence of relatively large offsets between the $I_\text{W24}$ and $I_\text{W08}$ profiles in the outer regions, which can be clearly seen for NGC 2903, NGC 3521, NGC 5055, and NGC 5457.
It echoes our finding in the mock tests (Appendix~\ref{app:comb}) that, using the residual-rescaled images or the convolved-model images as the interferometry data, produce similar $\SHI$ profiles after combining with the single-dish images. 
It highlights the powerful role of high-quality single-dish data in correcting for short-spacing, after which systematical uncertainties due to CLEAN residuals in the interferometric data are suppressed. 

In the following, we focus on $I_\text{W08}$ and $C_\text{W24}$ when comparing between THINGS-only and THINGS$+$FEASTS combined results, as they most clearly demonstrate the change from the work of \citetalias{Bigiel10}.
For most of the ten galaxies with FEASTS-combined $\hi$ data, the $C_\text{W24}$ $\SHI$ profiles systematically exceed the $I_\text{W08}$ profiles when the radius increases. 
The increments are insignificant in NGC 2841 and NGC 3198, which are also the galaxies with the smallest $\hi$ disks along the minor axes and nearly zero level of missing fluxes as found in W24. 
The increasing of fractional excess $\hi$ detected by FEASTS with radius has been demonstrated in W24, but at the relatively low resolution of 3$'$ of the FEASTS data. 
Here, after combining with the interferometric data, the resolution of profiles has been significantly improved, which is clearly illustrated as the significant difference between the $C_\text{W24}$ profiles and the FEASTS profiles. 

Figure~\ref{fig:prof2} compares the radial profile shapes for the subset of galaxies with new $\hi$ images from \citet{deBlok18, Koribalski18, Eibensteiner23} but without FEASTS data (see Section~\ref{sec:sample}). 
Like combining with single-dish images of FEASTS, combining with GBT images for well resolved galaxies or imaging with a compact array configuration also has helped to recover much more fluxes in these galaxies than the $I_\text{W08}$ data. 

The new $\SHI$ profiles ($C_\text{W24}$ profiles and the new profiles in Figure~\ref{fig:prof2}) exceed the THINGS-only profiles by an average of 0.08 dex at around $r_{25}$.
The average excess is 0.14 dex (0.28) dex at $R_{\rm HI, Iw08}$ ($R_{\rm HI, new}$), where $R_{\rm HI, Iw08}$ ($R_{\rm HI, new}$) is the radius where an $I_\text{W08}$ (new) profile $\SHI$ reaches 1 $\Msunpcsq$. 
If the whole sample is considered, the updated $\hi$ images increase $\SHI$ by an average of 0.06, 0.11, and 0.21 dex at these characteristic radii in comparison to the THINGS-only $I_\text{W08}$ data.

\subsubsection{The Pixelwise $\SHI$ Distributions}

\begin{figure} 
\centering
\includegraphics[width=8cm]{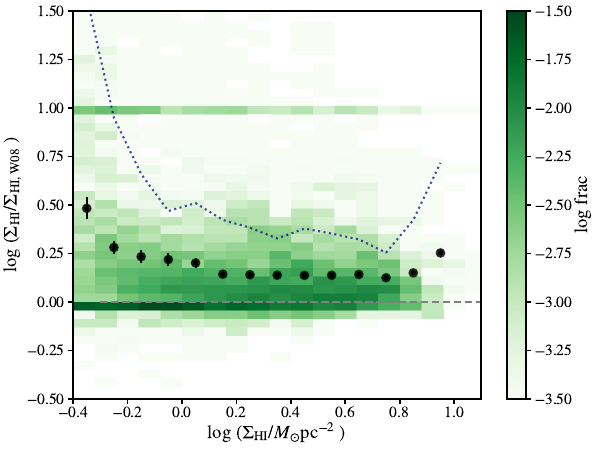}

\caption{ The difference in pixelwise $\SHI$ measurements between the $C_\text{W24}$ and $I_\text{W08}$ data in outer disks. 
The ratios of the two (former divided by the later, $\Sigma_{\rm HI}/\Sigma_{\rm HI,W08}$) are plotted as a function of the $C_\text{W24}$ measurements $\SHI$. 
The green color scales show the relative abundance of data points in the parameter space. 
The black dots and error bars show the mean differences and their 1-$\sigma$ errors of $\log \Sigma_{\rm HI}/\Sigma_{\rm HI,W08}$ as a function of $\SHI$. 
The grey dashed horizontal line mark the y position of zero. 
The blue dotted line shows the 90 percentile of $\Sigma_{\rm HI}/\Sigma_{\rm HI,W08}$ distribution at a given $\SHI$.
}
\label{fig:difSHI}
\end{figure}

Figure~\ref{fig:difSHI} shows the distribution of change in $\SHI$ with respect to the THINGS-only data due to updating $\hi$ images.
The $\SHI$ is measured from 1-kpc pixels in outer disks, and the distribution of change in $\SHI$ is plotted as a function of $\SHI$.
The $\SHI$ on average increases by $\sim0.15$ dex when $\SHI$ ranges from 1 to 5 $\Msunpcsq$. 

\subsection{Updated $\Ssfr$-$\SHI$ Relations}
\label{sec:SFL}
\begin{figure*} 
\centering
\includegraphics[width=8cm]{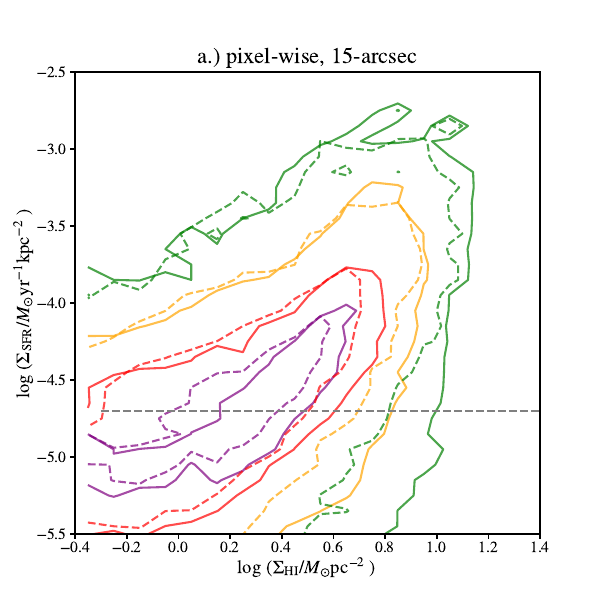}
\includegraphics[width=8cm]{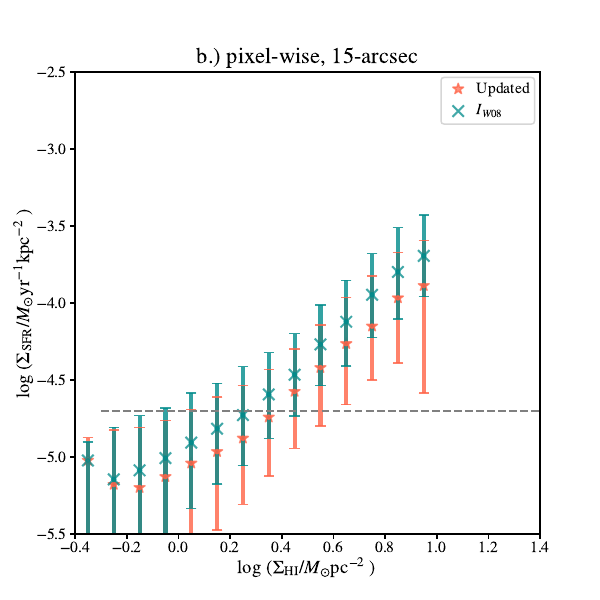}

\includegraphics[width=8cm]{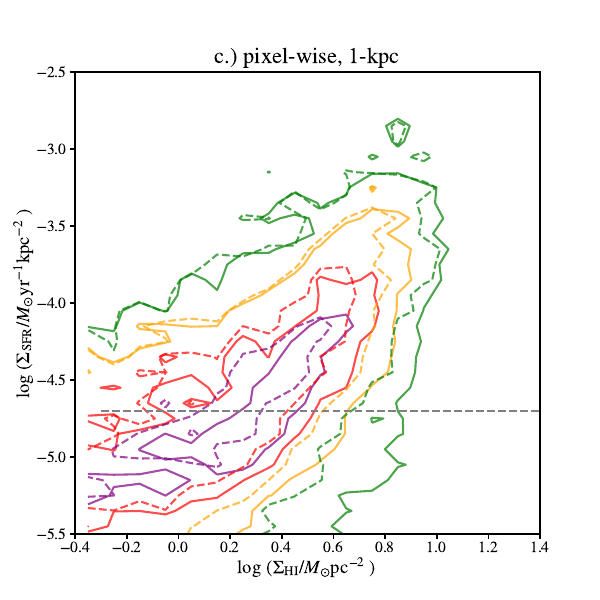}
\includegraphics[width=8cm]{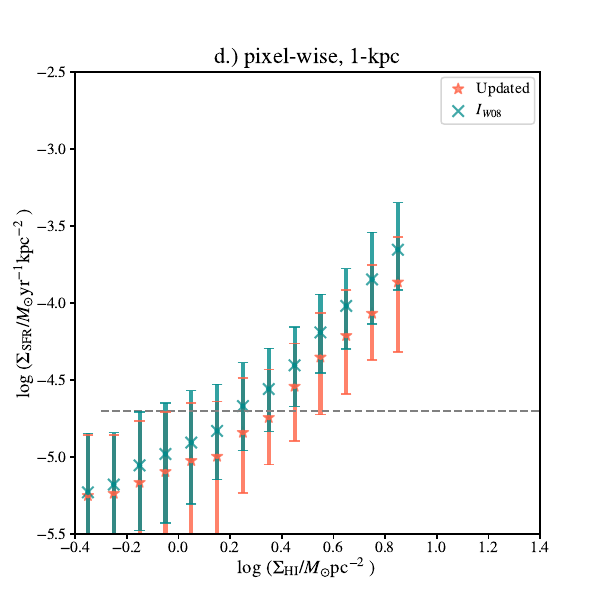}
\caption{ Pixelwise $\hi$-SFLs. 
{\bf Panel a:} distribution of pixelwise measurements in the space of $\Ssfr$ versus $\SHI$.
The result is based on images smoothed to the 15-arcsec resolution, and the plot is plotted in a similar way as Figure 8 in B10. 
The purple, red, orange, and green contours enclose 25, 50, 75, and 90 percents of the data points. 
The solid and dashed contours are for measurements based on the $C_\text{W24}$ and $I_\text{W08}$ $\hi$ images, respectively. 
The horizontal, grey dashed line at $\log (\Ssfr/\Msunyrkpcsq)=-4.7$ marks the typical 1-$\sigma$ uncertainty for pixelwise $\Ssfr$ measurements (section~\ref{sec:Sigma_measure}).  
{\bf Panel b:} the median relation of $\Ssfr$ as a function of $\SHI$. 
It is based on the same data plotted in panel a, but plots the median values and scatters of $\Ssfr$ in bins of $\SHI$ with 0.1 dex widths.
The red stars and cyan crosses plot the median values of $\Ssfr$ in $\SHI$ bins, for the $C_\text{W24}$ and $I_\text{W08}$ measurements respectively. 
The lengths of upper (bottom) error bars are calculated as 1$/$1.65 times the difference between the 90 (10) percentile and the median, which is roughly 1-$\sigma$. 
{\bf Panels c and d:} similar as panels a and b respectively, but the measurements are made from images smoothed to the 1-kpc resolution.
  }
\label{fig:SFL}
\end{figure*}

We show the pixelwise $\Ssfr$-$\SHI$ relation of outer disks in Figure~\ref{fig:SFL}, plotted in a similar way as Figure 8 in B10. 
The dashed line at $\log (\Ssfr/\Msunyrkpcsq)=-4.7$ marks the typical 1-$\sigma$ uncertainty for individual pixelwise $\Ssfr$ measurements. 

In panel a, we can see that with the $C_\text{W24}$ measurements, $\Ssfr$ distribution at a given $\SHI$ systematically shift downward as a result of systematically higher $\SHI$, in comparison to the distribution with the $I_\text{W24}$ measurements.
The downward shift is particularly clear for the lower half of the $\Ssfr$ distribution at a given $\SHI$.
The lower sides of contours of the $C_\text{W24}$ measurements at all levels are lower than the $I_\text{W08}$ measurements. 
At $\log (\SHI/\Msunpcsq)=0.8$ there is almost no data points with $\log (\Ssfr/\Msunyrkpcsq)$ below the $-4.7$ line in the $I_\text{W08}$ dataset, while in the $C_\text{W08}$ dataset, the $\log (\Ssfr/\Msunyrkpcsq)$ values extend all the way to the lower limit of the displaying range. 
On the other hand, except for the inner most 25-percentile contour, the upper sides of contours for the two types of measurements are close. 
The shift in median value and lower envelope of $\Ssfr$ at a given $\SHI$ is more directly quantified in panel b.
The average downward shift in median $\log \Ssfr$ is 0.16$\pm$0.03 dex at a fixed $\log (\SHI/\Msunpcsq)$ of $>0.3$, when we change from $I_\text{W08}$ to $C_\text{W24}$ measurements. 
The average lower error bars of $\log \Ssfr$ in the same $\SHI$ range are 0.28 and 0.43 dex, for the $I_\text{W08}$ and $C_\text{W24}$ measurements, respectively. 

Similar shift in median values and lower envelopes can be seen in the bottom row of Figure~\ref{fig:SFL}, when the measurements are based on images smoothed to the 1-kpc resolution. 
The average shift in median $\log \Ssfr$ is 0.18$\pm$0.03dex, and the average lower error bars for the $I_\text{W08}$ and $C_\text{W24}$ datasets are 0.27 and 0.36 dex, respectively.

\begin{figure*} 
\centering
\includegraphics[width=8cm]{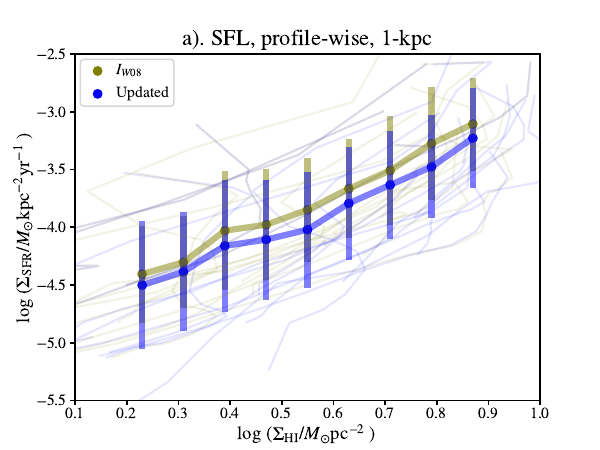}
\includegraphics[width=8cm]{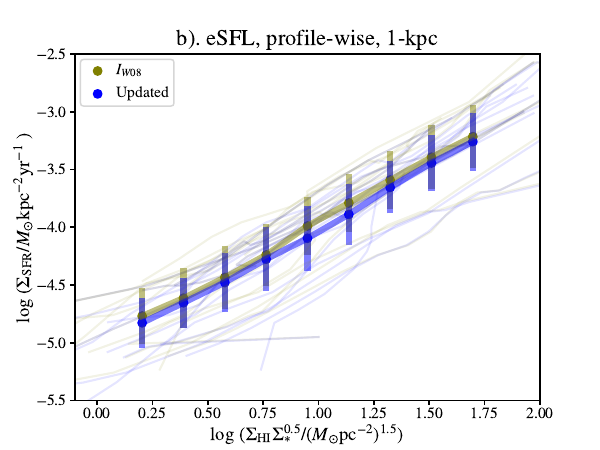}
\caption{ Azimuthally averaged relations of $\Ssfr$ and $\SHI$ in outer disks. 
Panel a plots the $\Ssfr$-$\SHI$ relation, and panel b plots the relation between $\Ssfr$ and $\SHI \Sigma_*^{0.5}$ which is often referred to as the extended SFL \citep{Shi11}.
The yellow and blue colors are for measurements based on the $I_\text{W08}$ and $C_\text{W24}$ images respectively. 
Each thin curve is for one galaxy.
The thick curve with dots and error bars show the mean relation and standard deviation of distribution.
}
\label{fig:SFL_rprof}
\end{figure*}

The azimuthally averaged version of the $\Ssfr$-$\SHI$ relation among the sample is presented in panel a of Figure~\ref{fig:SFL_rprof}, based on the 1-kpc resolution image set.
There is a similar downward shift in median relation by 0.15$\pm$0.03 dex due to the rightward shift in $\SHI$, when $\log (\SHI/\Msunpcsq)>0.3$. 

In total, the updated higher $\SHI$ values have led to a $\gtrsim$0.15 dex downward systematic shift in $\Ssfr$, together with a much larger lower-half scatter in $\Ssfr$ at a given $\SHI$ than the results previously based on the W08 dataset.

\subsection{Updated $\Ssfr$-$\SHI$-$\Sst$ Relations}
\label{sec:exSFL}
\begin{figure*} 
\centering
\includegraphics[width=8.5cm]{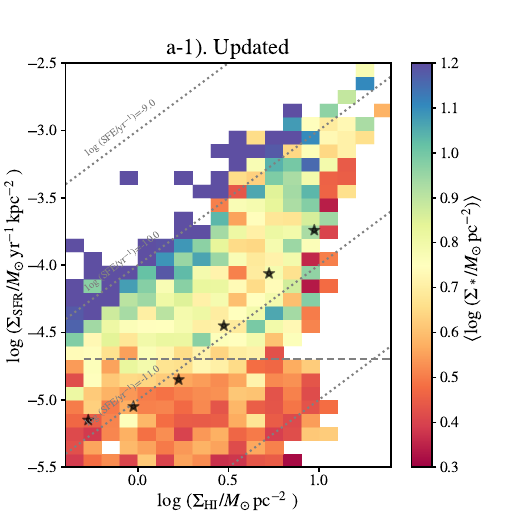}
\includegraphics[width=8.5cm]{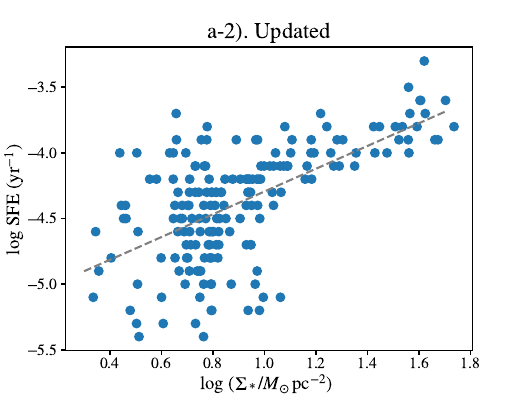}
\includegraphics[width=8.5cm]{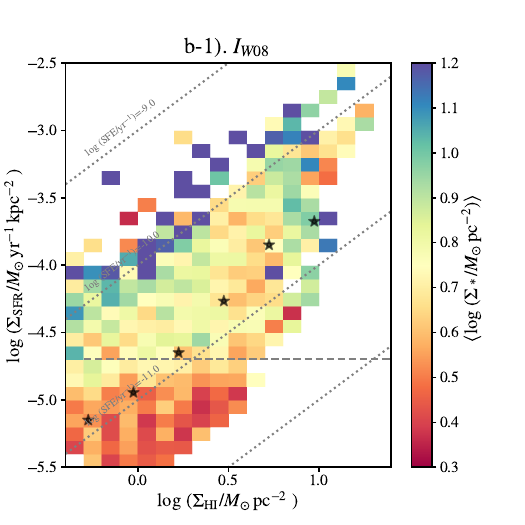}
\includegraphics[width=8.5cm]{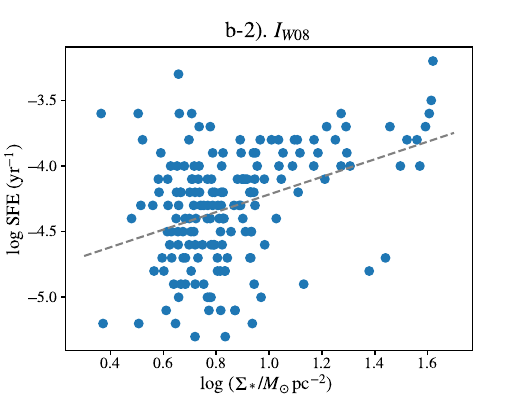}
\caption{ The distribution of average $\log \Sigma_*$ in the space of $\log \Ssfr$ versus $\log \SHI$. 
Panels labeled with a and b are for $C_\text{W24}$ and $I_\text{W08}$ measurements respectively. 
In each panel on the left column, the pixels are color coded by the average $\log \Sigma_*$ in bins that is 0.1 dex wide in both $\Ssfr$ and $\SHI$. 
The black stars show median $\Ssfr$ in bins of $\SHI$ that is 0.3 dex wide each.
The horizontal, grey dashed line at $\log (\Ssfr/\Msunyrkpcsq)=-4.7$ marks the typical 1-$\sigma$ uncertainty for pixelwise $\Ssfr$ measurements. 
The tilted, grey dotted lines mark positions of equal SFE with steps of 1 dex. 
In each panel on the right column, the relation between log SFE$=\log \Ssfr/\SHI$ is plotted as a function of $\log \Sigma_*$. 
The values are the average taken from the pixels above the $\Ssfr$ threshold in the corresponding plot on the left.
The best-fit linear relations are plotted in dashed lines and present in Section~\ref{sec:exSFL}.
}
\label{fig:eSFL_pixel}
\end{figure*}

In star formation theories, old stars can promote the formation of molecular gas that more directly form stars. 
The old stars contribute to the gravity increasing the dynamic pressure to accelerate gas collapse, and enrich the ISM with metals shielding the UV photons \citep{Ostriker10, Krumholz13}. 
In outer disks, they further help weaken strong perturbation effects and reduce star-formation stochasticities.
Thus, $\Sst$ is considered and found to be an important factor regulating the relation between $\Ssfr$ and gas surface densities.

The left column of Figure~\ref{fig:eSFL_pixel} shows the distribution of average $\Sigma_*$ in the space of $\Ssfr$ versus $\SHI$. 
The panels a-1 and b-1 are for the $C_\text{W24}$ and $I_\text{W08}$ measurements, respectively. 
We see in both panels that at a given $\SHI$, the $\Sigma_*$ increases with $\Ssfr$, which is consistent with the trend previously described by the extended SFL \citep{Shi18}.
Another way of describing the trend is that the star forming efficiency (SFE$=\Ssfr/\SHI$) increases with $\Sigma_*$, for the averege $\Sigma_*$ increases almost in parallel to the equal-SFE lines in the figure.

Though similar trends of SFE with $\Sigma_*$ are seen in both datasets, the trend is stronger in the $C_\text{W24}$ data, particularly at the high-SFE end. 
It is possible that the $I_\text{W08}$ under-estimated $\SHI$ (thus over-estimated SFE) artificially moved pixels with low $\Sigma_*$ upward, leading to an increased scatter in $\Sigma_*$ at a given position in the space of $\Ssfr$ versus $\SHI$.
The significant difference in $\Sigma_*$ between $C_\text{W24}$ and $I_\text{W08}$ measurements at the high SFE end is in contrast with the similar upper envelop in the distribution of pixelwise $\Ssfr$ at a given $\SHI$ shown in Figure~\ref{fig:SFL}. 

In order to demonstrate the trends and differences between two datasets more clearly, in the right column of  Figure~\ref{fig:eSFL_pixel}, we directly plot SFE as a function of $\Sigma_*$ by taking the average of values in pixels from the left plots.
The relations are clearest when $\Sigma_*>10~\Msunpcsq$, and have large scatter on the low-$\Sigma_*$ end, implying a surface density threshold for the old and intermediate-age stars to be at work.
When the $C_\text{W24}$ measurements are used, the Pearson and Spearman correlation coefficients for the relation of SFE with $\Sigma_*$ are 0.72 and 0.76, respectively, for the whole distribution of data.
The two coefficients reduce slightly to 0.62 and 0.56 when the data is limited to $\log (\Ssfr/\Msunyrkpcsq)>-4.7$.
These coefficients decrease to 0.66 (0.38) and 0.69 (0.34) when the $I_\text{W08}$ measurements are used instead for the whole parameter space ($\log (\Ssfr/\Msunyrkpcsq)>-4.7$ part). 
The updated $\SHI$ measurements helped to reveal a stronger dependence of SFE on $\Sigma_*$ than previously known with the W08 images. 

Linear regression is conducted with the $C_\text{W24}$ measurements for the relation of
\begin{equation}
\log ({\rm SFE/yr^{-1}})=a_0+ a_1 \log(\Sigma_*/\Msunpcsq) 
\label{eq:esfl}
\end{equation}
For the $\log (\Ssfr/\Msunyrkpcsq)>-4.7$ part of data, we obtain $a_0=-5.15\pm0.15$ and $a_1=0.85\pm0.16$. 
In comparison, if the regression is based on the $I_\text{W08}$ data, the slope $a_1$ would be a smaller value of 0.68.
The scatter of the best-fit relation is 0.34 dex, in contrast to a larger scatter of 0.40 dex if the linear regression is conducted with the $I_\text{W08}$ measurements.
The slope of 0.85 is much steeper than 0.5 from previous theoretical predictions \citep{Ostriker10} and observational deviations \citep{Leroy08} for inner disks of spiral galaxies, and observational deviations for dwarf galaxies \citep{Shi18}. 
Steepening in $\hi$-dominated outer disks of spiral galaxies was noticed before in \citet{Shi11}, but based on the W08 $\hi$ data (i.e. similar to results based on the $I_\text{W08}$ measurements), so the extent is much smaller.

In order to examine any residual dependence on $\SHI$ in Equation~\ref{eq:esfl}, we conduct linear regression for the following relation, treating the $\SHI$ as a variable:
\begin{equation}
\log \Ssfr=b_0+ b_1 \log \SHI + b_2 \log \Sst ,
\label{eq:esfl2}
\end{equation}
where $\Ssfr$, $\SHI$, and $\Sst$ has the unit of $(\Msunyrkpcsq)$,  $(\Msunpcsq)$, and $(\Msunpcsq)$, respectively.
For the $\log (\Ssfr/\Msunyrkpcsq)>-4.7$ part of data, we obtain $b_0=-5.13\pm0.11$, $b_1=0.98\pm0.07$, $b_2=0.85\pm0.09$, and scatter of 0.34 dex.
The $b_1$ does not significantly deviate from unity, supporting the treatment of combining $\Ssfr$ and $\SHI$ into SFE in Equation~\ref{eq:esfl}. 

We notice that, the previous extended SFLs in the literature tended to rely on azimuthally averaged instead of pixelwise measurements.
We also use the azimuthally averaged measurements, and plot the relations of $\Ssfr$ versus $\SHI \Sigma_*^{0.5}$ in panel b of Figure~\ref{fig:SFL_rprof}.
The $I_\text{W08}$ and $C_\text{W24}$ measurements lead to close median relations, with an average offset of 0.07$\pm$0.03 dex along the y direction.
Both relations have a 1-$\sigma$ scatter of 0.26 dex. 
The significantly reduced offset and scatter compared to the $\Ssfr$ versus $\SHI$ relation (Figure~\ref{fig:SFL_rprof}-a) imply that $\Sigma_*$ also plays an important role in determining SFR in these outer disks. 
The significantly reduced offset and scatter compared to the pixelwise relation (Figure~\ref{fig:SFL}-a) are likely because the azimuthally average measure is biased toward the high surface density values at the same radii, where the SFL is less sensitive to the fractional change in $\SHI$.

\section{Discussion}
\label{sec:discussion}
\subsection{Possible Influence of Internal Dust Attenuation}
We apply a simple correction for the internal dust attenuation of FUV luminosities, assuming $\text{E}(\text{B}-\text{V})=N_{\rm HI}/(5.8\times10^{21}~{\rm cm}^{-2})$ mag \citep{Bohlin78} and A$_{\rm FUV}= 8.24 \text{E}(\text{B}-\text{V})$ \citep{Wyder07}.
We note that such a correction may be over-simplified, as there should be a metallicity dependence in the ratio $\text{E}(\text{B}-\text{V})/N_{\rm HI}$.
After the dust attenuation correction, $a_0=-4.90\pm0.21$ and $a_1=0.87\pm0.23$ in Equation~\ref{eq:esfl}, and the scatter increases significantly to 0.52 dex. 
The increase in scatter may not be so surprising, since the adopted way of attenuation correction has the mathematical effect of  steepening the $\Sst$ contours in Figure~\ref{fig:eSFL_pixel}, making them less parallel to the equal-SFE lines.
Thus, Equation~\ref{eq:esfl2} may be the better characterization of the relation between $\Ssfr$, $\SHI$, and $\Sst$, for which we obtain $b_0=-5.58\pm0.14$, $b_1=1.62\pm0.10$, and $b_2=1.26\pm0.12$. 
The coefficients $b_1$ and $b_2$ are much larger than unity and $a_1=0.87$, respectively, but the scatter of the relation is 0.49 dex, not much lower than that of Equation~\ref{eq:esfl}.


Either the A$_{\rm FUV}$ has introduced in large uncertainties due to possibly steep gradients in metallicity distributions in outer disks of some galaxies \citep{Moran12}, or the intrinsic scatter of the extended SFL is indeed so large, reflecting physics not captured by Equation~\ref{eq:esfl2}. 
This caveat should be investigated in the future with spectroscopic data to more directly constrain the metallicity and dust attenuation. 
Our following discussion is based on results obtained in Section~\ref{sec:result}.

\subsection{The Size--Mass Relation and Disk Confinement of HI}

\begin{figure} 
\centering
\includegraphics[width=8cm]{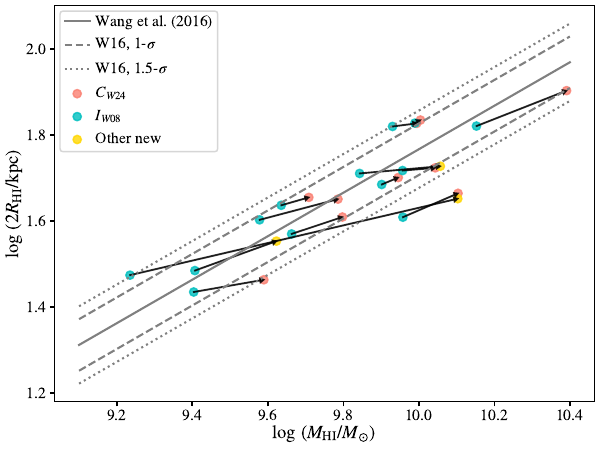}
\caption{  The $\hi$ size-mass relation. 
The red, green, and yellow dots are for measurements made with the $C_\text{W24}$ images, the $I_\text{W08}$ images, and new $\hi$ images published in previous papers (see section~\ref{sec:data}).  
Each grey arrow links measurements from different $\hi$ images for the same galaxy. 
The black solid, dashed, and dotted lines mark the size-mass relation of \citet{Wang16}, its 1-$\sigma$ deviation, and its 1.5-$\sigma$ deviation, respectively.  }
\label{fig:MD}
\end{figure}

Table~\ref{tab:sample} lists the $\hi$ mass and radius measurements based on $I_\text{W08}$ and new data.
Previously, based on interferometric $\hi$ images, it was found that all galaxies lie tightly on a $R_{\rm HI}$-$M_{\rm HI}$ relation, with a 1-$\sigma$ scatter of only 0.06 dex along the $R_{\rm HI}$ axis \citep{Wang16}. 
Since both $\hi$ masses and $R_{\rm HI}$ change due to the use of new $\hi$ images, the positions of galaxies are expected to shift in the $\hi$ size-mass diagram, which is confirmed in Figure~\ref{fig:MD}.  
The data points on average shift right-ward and the trend becomes shallower, because fractionally more excess $\hi$ is detected at large radius.
Out of 13 galaxies with newly measures $R_{\rm HI}$, seven lie 1-$\sigma$ below the relation, while with the $I_\text{W08}$ measurement only one galaxy, NGC 5055, is 1-$\sigma$ below the relation.  

These seven outlier galaxies have an average $\log M_{\rm HI}/M_{\rm HI,W08}$ of 0.27 (with scatter 0.25), in comparison to the average of 0.10 (with scatter 0.08) for the rest sample.
The galaxies missing more $\hi$ fluxes in the interferometric data tend to be the ones having larger angular-size $\hi$ disks, and$\slash$or experiencing stronger tidal interactions \citepalias{Wang24}. 
Particularly, there are two galaxies (NGC 5194 and NGC 5236) having $I_\text{W08}$ and new measurements within and below the 1.5-$\sigma$ limit of the relation, respectively.  
NGC 5194 is the most strongly tidally interacting system among the sample, and NGC 5236 is known for its (possibly also tidally perturbed) extraordinarily extended $\hi$ and UV disks \citep{Thilker05, Koribalski18}, which may explain their extreme behavior around the relation. 

The interferometry-detected $\hi$ in the nearby universe tends to be clumpy, dense, and kinematically cold $\hi$ in a thin disk, while the interferometry-missed $\hi$ the smooth, diffuse and kinematically hot $\hi$ in a thick disk \citepalias{Wang24}.  
It is possible that only the dense $\hi$ that is relatively well confined to the disk mid-planes, is more strongly regulated by galactic internal physics, and thus tends to follow the the very tight size-mass relation \citep{Wang16}.
A further implication is that, once some of the $\hi$ tends to escape the disk confinement and regulation, it firstly go through the diffuse $\hi$ phase.
We will come back to the size-mass relation topic in a future study with a larger and more diverse sample.

\subsection{The Role of Interferometry-missed HI in Star Formation}
We have shown that after short-spacing correction, the median $\hi$ SFL for the same B10 sample shifts down-ward moderately ($\gtrsim$0.15 dex) and its scatter also increases moderately ($\sim$0.15 dex), while the SFE-$\Sigma_*$ correlation and relation have strengthened and tightened significantly.  
The updating of these relations is relatively straightforward, given the systematic increase of $\SHI$ with the new $\hi$ observations. 
A question is, whether the changes in these relations and correlations due to the inclusion of diffuse $\hi$ are physically meaningful.
After all, the thermodynamically colder dense $\hi$ may be more closely linked to the final star formation  \citep{Hennebelle19, Krumholz18, Ostriker22}. 
In theory, gas needs to go through cooling, collapsing, and fragmenting into dense clouds and filaments, before forming stars \citep{Hennebelle19, Krumholz18, Ostriker22}. 
The role of diffuse and total $\hi$ in star formation can thus be appreciated from at least the following aspects.

Firstly, the diffuse gas may be replenishing reservoir for the dense star-forming gas. 
The molecular clouds where star formation takes place will be exhausted within $\sim$2 Gyr and should be replenished for maintaining the star formation activity over the timescale of galaxy evolution \citep{Bigiel08}.
The cold dense $\hi$ is supposed to be the indispensable reservoir and the ingredient of molecular clouds \citep{Inoue12, Inutsuka15}.
Possibly because of this connection, observationally, when the volume densities are considered, the interferometry-detected dense $\hi$ has similarly tight relation to the SFR as the molecular gas \citep{Bacchini19,Bacchini19b, Bacchini20}.
The diffuse HI disk is likely a thicker envelop of the dense HI gas \citepalias{Wang24}, and thus possibly replenishes the dense $\hi$ reservoir. 
In the outer disks where both star formation and SN feedback are clearly inefficient, gas collapse may even be able to start from relatively diffuse phases without being disrupted by feedback \citep{Elmegreen18}. 
In the future, it will be interesting to derive volume densities of the total and diffuse $\hi$ for the spiral galaxies, and revisit the volumetric SFL. 

Secondly, it provides the necessary condition for star formation in a self-regulated ISM \citep{Ostriker22}.
The diffuse gas (both $\hi$ and molecular) possibly contributes to both the ambient pressure to confine the molecular clouds \citep{Sun20}, and the turbulence to keep molecular clouds from catastrophic collapse \citep{Hennebelle06}. 
It is thus important to detect the total $\hi$ for a full budget of mass, pressure, feedback energy, and turbulence.
And it may be useful to further separate the diffuse and dense $\hi$ to dissect the self-regulation picture.
Radial profiles of $\SHI$ (Section~\ref{sec:profile}) and velocity dispersion \citepalias{Wang24} support the possibly important contribution of diffuse $\hi$ to the weight and turbulence of star-forming regions in outer disks. 

Thirdly, it carries possible clues to understanding the large scatter of the SFL in low-$\hi$-column-density regimes.
The scatter may reflect regulation by localized environment (e.g., the pressure, \citealt{Ostriker10}), or different gas-instability--driving mechanisms (e.g. local dynamics versus feedbacks, \citealt{Semenov18}), or different temporal stages of the star formation cycle (e.g., the lag between cloud assembly and star ignition, and between star formation and feedback, \citealt{Orr19}). 
Correctly measuring the properties of total and diffuse $\hi$ is an important first step toward handling these complexities, which is supported by the strengthened correlation between SFE and $\Sigma_*$ with updated $\SHI$ (Section~\ref{sec:exSFL}).
Possibly useful information carried by the diffuse $\hi$ kinematics awaits exploration.

Finally, it possibly traces the CGM-ISM connection, the influence of which on triggering star formation activities starts to raise interests in theoretical studies \citep{Gurvich23}. 
As the galactic fountain modifies the amount, kinematics, and metallicity of the ISM, it may directly influence the localized star formation process \citep{Shimoda24}.
While the stratification of CGM near galaxy disks is poorly constrained, the ambient pressure from the CGM propagating through the total $\hi$ (and firstly the diffuse $\hi$) may play a role in setting the pressure of the star-forming gas. 
As the diffuse $\hi$ is closer to the CGM than the dense $\hi$, its properties are possible diagnostics for these CGM effects. 

Because of these aspects, fully capturing the total as well as the diffuse $\hi$ has the potential to advance our understanding of star formation.
It is possibly also a necessary step toward fully linking localized star formation to the internal environment and dynamic evolution of galaxies.

\subsection{Implication for Star Formation and Galaxy Evolution}
Star formation models for $\hi$-dominated regions are often calibrated against THINGS measurements \citep{Ostriker10, Krumholz13}, which are further implemented as subgrid physical models in cosmological simulations of galaxy formation. 

The changes in SFLs presented in this paper imply quantitative adjustment in the parameters or interpretation of previous SF models.
The systematic downward $\gtrsim0.15$ dex shift in $\Ssfr$ at a given $\SHI$ indicates an even more inefficient SFE than previously captured. 
The increased scatter in the $\Ssfr$-$\SHI$ relation emphasizes the larger diversity in conditions for star formation. 
The strengthened dependence of SFE on $\Sigma_*$ highlights the elevated importance of evolved stars in regulating SFE in outer disks. 
The role of these evolved stars, as suggested by previous models, can be possibly through a combined effect of providing dynamic pressure \citep{Ostriker10} and metallicity \citep{Krumholz13}. 
They promote the formation of molecular gas, modulate feedback strengths, restore the gas against violent perturbations, and reduce the stochasticity of star formation \citep{Semenov18, Orr19}. 

The strengthened dependence of SFE on $\Sigma_*$ may be consistent with the slope of dependence being steeper than 0.5. 
Following standard SF models (e.g., \citealt{Krumholz13}), the star-forming process in cloud scales can be characterized as 
\begin{equation}
\Ssfr=\epsilon \frac{ \Sigma_{\rm gas} }{t_{\rm ff} },
\end{equation}
where $\epsilon$ is the cloud-scale star-forming efficiency.
In the inner disk, the stars dominate the gravity, and the mid-plane pressure is proportional to $\Sigma_*^{0.5}$, leading to $t_{\rm ff}\propto\Sigma_*^{-0.5}$, well explaining the dependence of SFE on $\Sigma_*^{0.5}$ there \citep{Ostriker10}.
The steeper dependence of SFE on $\Sigma_*^{0.5}$ on outer disks implies that, (1) the mid-plane pressure is no longer proportional to $\Sigma_*^{0.5}$, as the stellar disk become less important in gravity in presence of gas and dark matter, and possibly flares out; (2) $\Sigma_*$ possibly affects $\epsilon$, through type Ia SN and AGB feedback, as the stellar population in these UV-bright outer disks is relatively younger than in the inner disk. 
These intermediate-age stellar products enrich the ISM with metals that enhance gas cooling and molecular formation \citep{Krumholz13}, and SN driven shocks may compress the warm gas leading to formation of cool gas and stars \citep{Ostriker22}.
Both possible reasons highlight the influence of stellar disks on SF conditions.

Cosmological galaxy simulations are typically calibrated against observed stellar mass function and optical correlation function in different redshift intervals \citep{Somerville15, Naab17}, and as a result the average star forming histories of galaxies are likely not much influenced by a moderate change in SFL.
The change in SFL in $\hi$-dominated regions may propagate back to affect more the gas properties at each redshift, particularly in low-mass ($M_*\lesssim10^{9}\, M_{\odot}$) galaxies and galaxy outer disks. 
Around 44\% of $\hi$ is contained in galaxies with $M_*<10^9\, M_{\odot}$ (estimated based on the NeutralUniverseMachine model of \citealt{Guo23}), so this change in SFL can significantly influence the predicted cosmic $\hi$ content.  
This effect is demonstrated in recent theoretical studies predicting signal of cosmic $\hi$ intensity mapping \citep{Wang21}.
The update in SFL of $\hi$-dominated regions can be conducive to updating$\slash$calibrating $\hi$-H$_2$ transition prescription in the simulation, particularly in regions of low gas densities. 
The prescription is necessary in realizing $\hi$ and H$_2$ partition in most cosmological simulations that only trace gas temperature down to $10^4\,$K  \citep{Crain23}. 
The predicted H$_2$ scaling relations in simulations may need to be modified in this sense. 
In addition to the SFLs, the systematic increase of $\SHI$ also impacts the dust$/\hi$-based measurement of the CO-to-H$_2$ conversion factor \citep{Sandstrom13}, whose absolute calibration often relies on the THINGS $\hi$ data. 
As molecular gas masses are often inferred from CO observations, the change in this conversion factor leads to changes in both $\hi$-H$_2$ and H$_2$-SFR prescriptions. 
The $\hi$-dominated low-mass galaxies and galaxy outer disks do not contribute much to the cosmic total budget of stellar mass and SFR, and thus do not evolve ``by construction'' of observational constraining in simulations. 
Their assembly history may change with the updated SFL. 
Galaxy hydro-dynamic simulations have difficulties in reproducing the neutral gas properties at relatively high redshift, as well as the CGM multi-phase structures \citep{Crain23}. 
They also do not reach a consensus yet on how to correctly reproduce kinematics of dwarf galaxies and disk-bulge fine structures of more massive galaxies \citep{Crain23}.
Capturing the total $\hi$ gas in its relation with the SFR, as part of the troubleshooting for baryonic physics, may be helpful in tackling these galaxy simulation challenges.

\subsection{Caution about Using Interferometry-only Data to Study SFLs}
The roles of diffuse and dense $\hi$ in star formation also indicate a few observational systematic effects that may worth our attention when studying SFLs using interferometric-only data, the size of which grow quickly with SKA pathfinder surveys \citep{Koribalski20}. 
The interferometry-missed $\hi$ flux ratio strongly depends on disk angular sizes (instead of physical size) and observational conditions (including array configuration, RFI contamination, and sensitivity) \citepalias{Wang24}. 
As a result, the interferometry detected $\hi$ fluxes can be either an arbitrary portion of the dense $\hi$, or a portion of the diffuse $\hi$ plus the dense $\hi$.
This effect may contribute in an non-physical way to the scatter of $\hi$ SFLs and other $\SHI$-related relations, particularly in $\hi$-dominated regions ($\gtrsim 0.4~r_{25}$, \citealt{Leroy08}).
Caution is thus needed when combining and comparing galaxies with different $\hi$ masses (due to the size-mass relation of $\hi$, \citealt{Wang16}) or at different redshifts.
Comparing the integrated fluxes of interferometry data with those from single-dish data is a helpful examination of the possible systematic uncertainty.

\section{Summary and Conclusion}
\label{sec:summary} 
We have updated a subset of the THINGS sample in its $\hi$ surface density measurements, which have been used extensively to study the SFL in galaxy outer disks (beyond the optical $r_{25}$) since the benchmark work of B10, based on new observations of FEASTS and supplemented with new data in the literature.
The update is to correct for or mitigate the short-spacing problem of previous VLA observations, by combining the VLA images with FEASTS images taken by FAST (or with $\hi$ images taken by GBT for the two galaxies with largest apparent sizes as done in the literature) when data are available. 
The $\hi$ surface densities increase by a median of 0.15 dex in the outer disks, and for individual galaxies the increments are larger toward larger radius. 
As a result the relation of pixelwise $\Ssfr$ versus $\SHI$ move downward by $\sim$0.15 dex, and the lower envelop of the scatters shifts downward even more. 
The scatters are found to strongly correlate with the $\Sigma_*$. 
The relation between $\Ssfr/\SHI$ and $\Sigma_*$ is much tighter and the slope steeper than previously based on the original THINGS $\hi$ images.

\section*{acknowledgments}
We thank S. Faber, C. Li, X. Kong for useful discussions, and thank the anonymous referee for very constructive comments. 
JW thanks support of the research grants from Ministry of Science and Technology of the People's Republic of China (NO. 2022YFA1602902),  the National Natural Science Foundation of China (NO. 12073002), and the science research grants from the China Manned Space Project (NO. CMS-CSST-2021-B02).  
LCH was supported by the National Natural Science Foundation of China (11721303, 11991052, 12011540375, 12233001), the National Key R\&D Program of China (2022YFF0503401), and the China Manned Space Project (CMS-CSST-2021-A04, CMS-CSST-2021-A06).
Parts of this research were supported by the Australian Research Council Centre of Excellence for All Sky Astrophysics in 3 Dimensions (ASTRO 3D), through project number CE170100013.
Parts of this research were supported by High-performance Computing Platform of Peking University.

This work made use of the data from FAST (Five-hundred-meter Aperture Spherical radio Telescope). FAST is a Chinese national mega-science facility, operated by National Astronomical Observatories, Chinese Academy of Sciences.

\facilities{FAST: 500 m, VLA }
\software{Astropy \citep{astropy:2013, astropy:2018, astropy:2022}, 
numpy \citep[v1.21.4]{vanderWalt11}, photutils \citep[v1.2.0]{Bradley19}, Python \citep[v3.9.13]{Perez07}, scipy \citep[1.8.0]{Virtanen20} }


\appendix   
\section{Tests on combining THINGS and FEASTS HI Images}
\label{app:comb}
\subsection{The Mock Data}
In W24, we have developed a procedure to generate mock VLA and FAST observations for simulated $\hi$ disks, for the purpose of examining uncertainties in cross-calibration of fluxes between the two types of data. 
The simulated $\hi$ disks have similar power spectra of $\hi$ column densities as the THINGS galaxies, with slope index $\kappa$ ranging from 1.8 to 3.2. 
They have $R_{\rm HI}=9'$, and follow the size-mass relation and average outer-disk radial profile shape of real galaxies \citep{Wang16}.
The mock FAST observations have the typical sensitivity and beam shape of FEASTS. 
The mock VLA observations mimic the THINGS observations, and are produced with the CASA task {\it simobserve}. 
The parameters of  {\it simobserve} are adjusted to produce a series of mock VLA observations with median SNR ranging from 0.9 to 4.4 in flux-detected regions of the final images. 
The mock VLA visibilities are reduced using the CASA task {\it tclean} with the multi-scale deconvolver, with a similar parameter setting as for the THINGS data reduction.
Three types of reduced VLA images for each mock observation are recorded, the standard image, the (residual) rescaled image, and the convolved model.

\subsection{Combining Different Types of VLA images with a FAST image}
We use the procedure described in Section~\ref{sec:comb} to combine the mock VLA and mock FAST images, with the goal of determining the best way of conducting the combination, and assessing possible systematic biases in the final products.
To be in line with the science analysis and goal of this paper, the diagnostics of image combination quality are how well the radial profiles and pixelwise surface densities are recovered. 
The image type and SNR of VLA images have been found to be major factors affecting the robustness of flux cross-calibration \citepalias{Wang24}, and they are thus expected to be also important in determining the image combination quality.
We will compare the combined images produced using the three types of reduced VLA images, and refer to them the standard-combined, the rescaled-combined, and the convolved-model-combined images, respectively. 
The power spectral slope determines the fraction of fluxes carried by small and large scale structures, thus may also affect the combination of images. 
The mock set which has a $\kappa=$2.2 and median SNR of 2.8 in the VLA data is taken as the reference set, for it is the closest to the typical condition of the THINGS data in the B10 sample.

\subsubsection{Test on Reproducing pixelwise HI Surface Densities}

\begin{figure} 
\centering
\includegraphics[width=9cm]{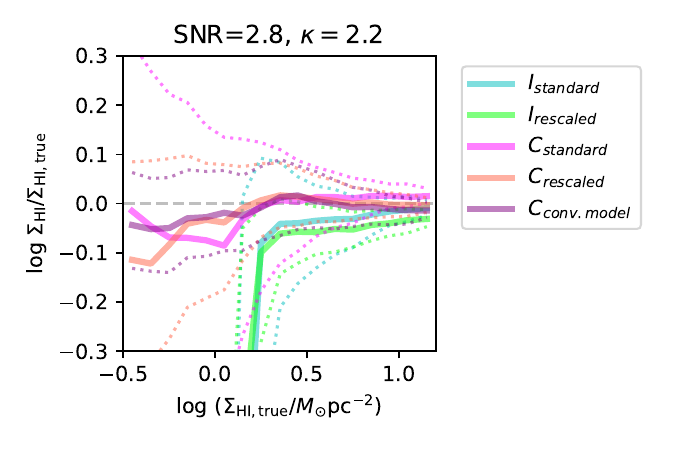}
\caption{
The recovering capabilities of pixelwise $\SHI$ of different types of $\hi$ images for the $\kappa=2.2$ \& SNR$=2.8$ mock dataset. 
The simulated disk based on which the mock images are taken has a power spectral index $\kappa=2.2$, and the VLA data has a median SNR$=2.8$.
Different colors show different types of mock $\hi$ images from which the measurements are made.
For each mock $\hi$ image type, the logarithm difference of $\SHI$ measured from mock images from the true values are plotted as a function of the true $\SHI$. 
The solid, thick curves show the median distribution of $\SHI$ difference as a function of true $\SHI$, and the dotted curves the scatters. 
The scatters are measured as the 25 and 75 percentiles. 
 }
\label{fig:pixNHI_mock_ref}
\end{figure}

Figure~\ref{fig:pixNHI_mock_ref} shows the pixelwise $\SHI$ measured from different images in comparison to the true image from the reference mock set. 
The combined images and VLA images reproduce the true $\SHI$ almost equally well when $\SHI \sim 10~ \Msunpcsq$, except that the rescaled VLA image and the standard-combined image under- and over-estimate slightly the true $\SHI$. 
When $\SHI$ is between 2 and 7 $\Msunpcsq$, the VLA images medianly deviate away from the true $\SHI$ by $\gtrsim0.05$ dex, while the combined images almost equally closely trace the true $\SHI$ on the median. 
But the scatter of the offset of the standard-combined image starts to become very large in this range.
Finally, when $\SHI<2~\Msunpcsq$, the rescaled-combined image has a larger median deviation from the true $\SHI$ than the convolved model-combined image, and also much larger scatter in the offset.
Throughout the $\SHI$ range from 0.5 to 20 $\Msunpcsq$, the convolved-model-combined image always has a median offset less than 0.05 dex, and a scatter $\lesssim$0.1 dex. 
On the whole, the combined images more closely recover the true $\SHI$ than the VLA-only images, and  the combined image that works best is the convolved-model-combined image. 

\begin{figure*} 
\centering
\includegraphics[width=8cm]{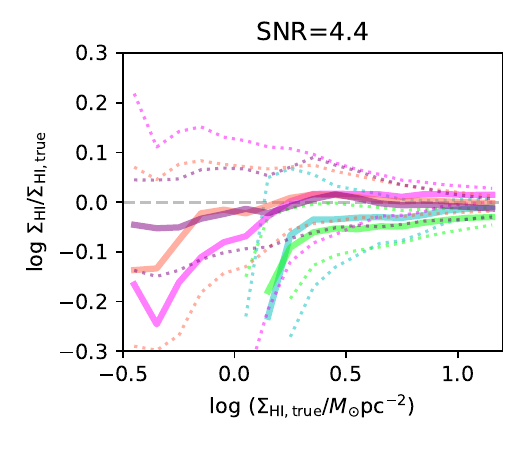}
\includegraphics[width=8cm]{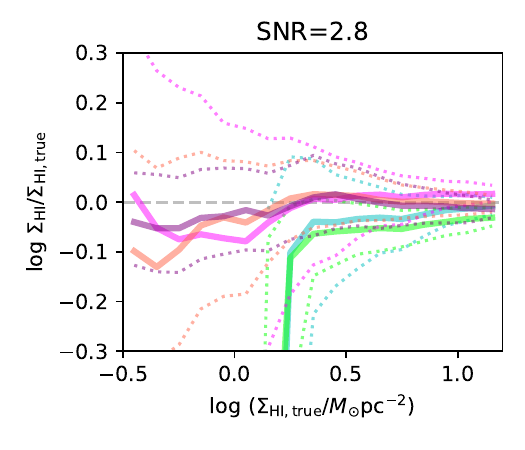}
\includegraphics[width=8cm]{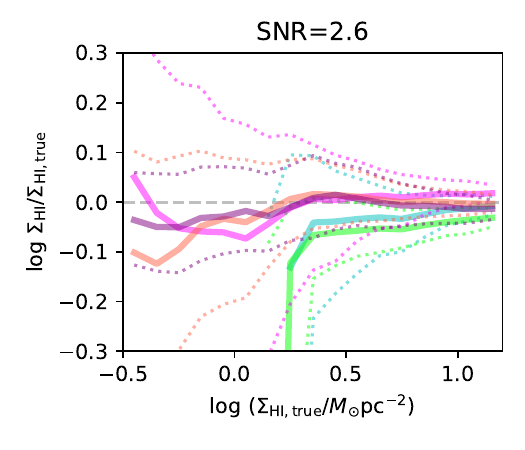}
\includegraphics[width=8cm]{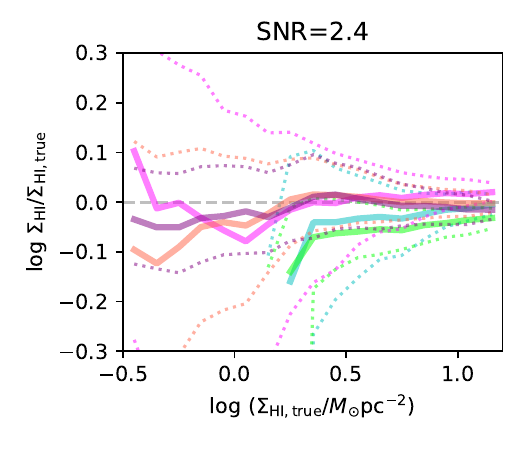}
\includegraphics[width=8cm]{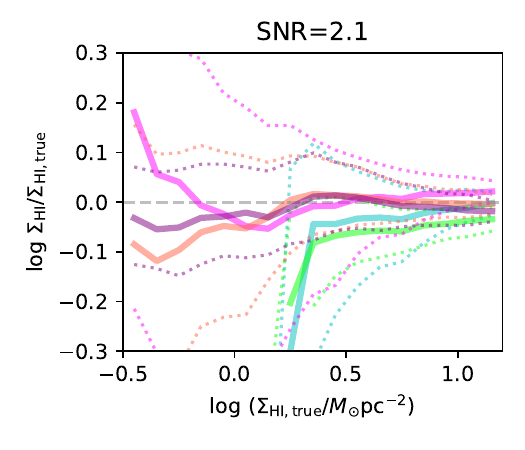}
\includegraphics[width=8cm]{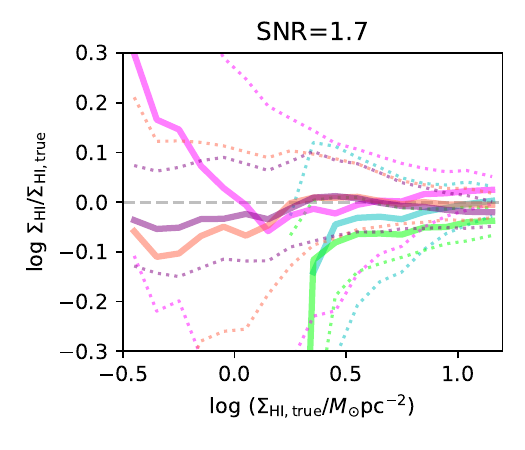}
\caption{ The recovering capabilities of pixelwise $\SHI$ of different types of $\hi$ images. 
Similar as Figure~\ref{fig:pixNHI_mock_ref}, but the VLA data have different SNR levels as labeled in each panel. 
 }
\label{fig:pixNHI_mock_snr}
\end{figure*}

\begin{figure*} 
\centering
\includegraphics[width=8cm]{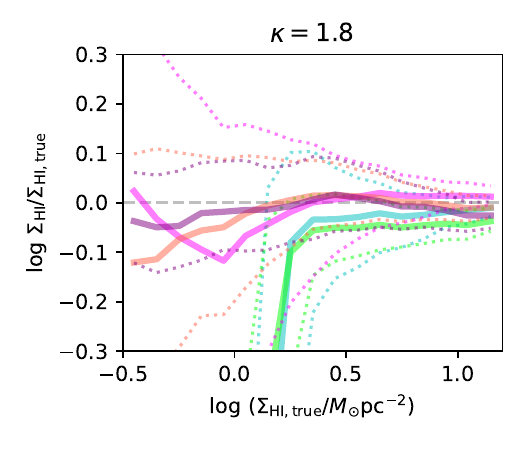}
\includegraphics[width=8cm]{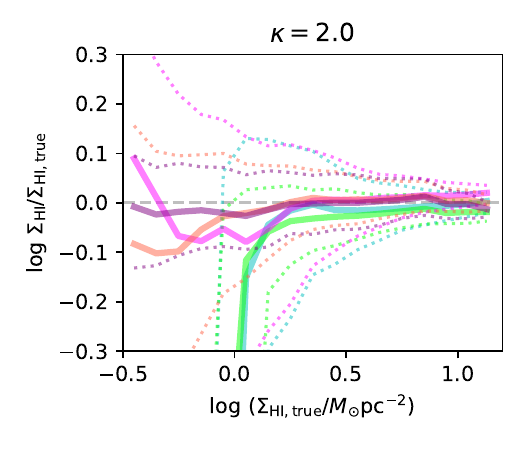}
\includegraphics[width=8cm]{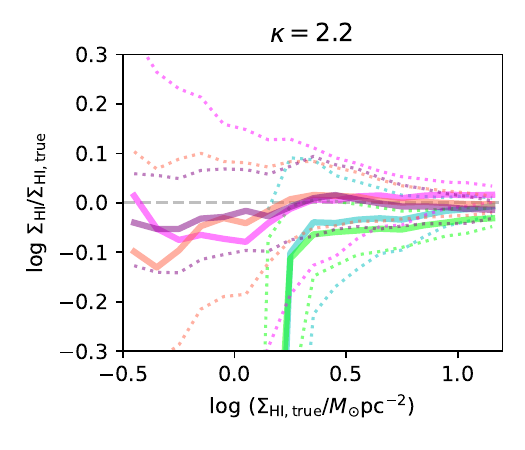}
\includegraphics[width=8cm]{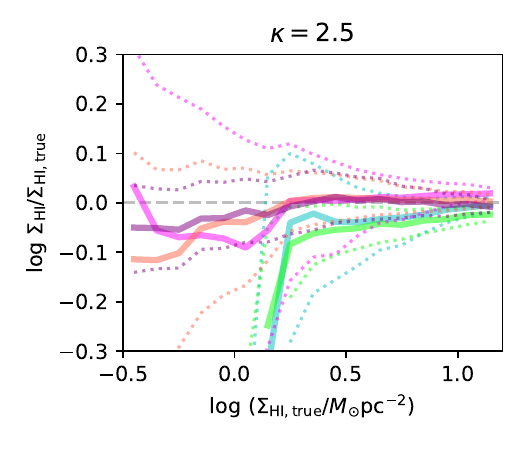}
\includegraphics[width=8cm]{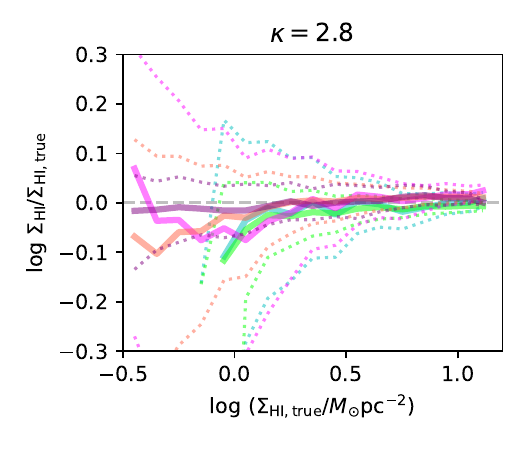}
\includegraphics[width=8cm]{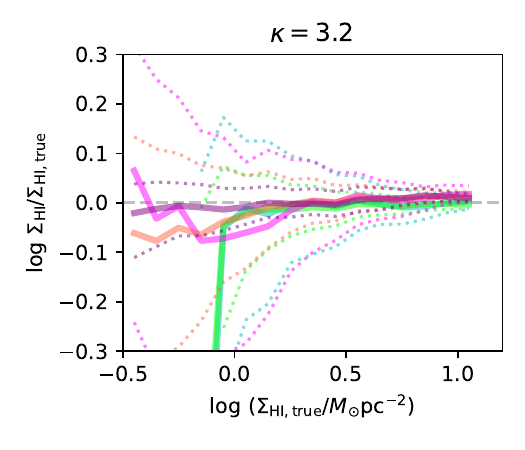}
\caption{  The recovering capabilities of pixelwise $\SHI$ of different types of $\hi$ images. 
Similar as Figure~\ref{fig:pixNHI_mock_ref}, but the mock disks have different power spectral indexes $\kappa$ as labeled in each panel.
}
\label{fig:pixNHI_mock_kappa}
\end{figure*}

Figure~\ref{fig:pixNHI_mock_snr} expands the test in Figure~\ref{fig:pixNHI_mock_ref} to mock images based on the same simulated disk but having different SNR for the VLA mock images. 
Figure~\ref{fig:pixNHI_mock_kappa} expands the test to simulated disks with different $\kappa$ values, but the SNR of VLA images are similar as in Figure~\ref{fig:pixNHI_mock_ref}. 
Similar conclusion can be reached that the convolved-model-combined image best recovers the true $\SHI$; other types of images have larger systematic offset on the median and$\slash$or larger scatter. 

It is also interesting to point out the following features. 
When $\text{SNR}\lesssim$2.4, systematic uncertainties of the flux cross-calibration factor between the VLA and FAST data rise quickly \citepalias{Wang24}, but the recovering of fluxes with combined images remains quite robust in the same SNR range. 
Possibly this is because the image combination (particularly in outer disk regions) in comparison to the flux cross-calibration benefits more from the high SNR of the FAST data. 
When $\kappa\gtrsim2.8$ and thus a high fraction of fluxes are contained in small-scale structures, the VLA images recover the true $\SHI$ very well before getting close to its detection limit. 
This type of galaxies is relatively rare, for example, consisting of only one tenth of the THINGS sample (Appendix D in W24).
In many cases, the VLA standard images seem to produce less median bias in $\SHI$ than the VLA rescaled images, but their scatters of offsets are much larger.

\subsubsection{Test on Reproducing Azimuthally Averaged HI  Surface Densities}
Figure~\ref{fig:pixNHI_mock_ref} shows the $\SHI$ radial profiles measured from different images in comparison to the true image for different mock datasets. 
Here the analyses of four representative mock datasets are displayed, which include the reference mock dataset, a mock dataset with similar $\kappa$ but lower SNR, and two mock datasets with similar SNR as the reference set but lower or higher $\kappa$ values.

The VLA image-based measurements are systematically lower than the true profile by $\gtrsim$0.05 dex when $r<500''$, except when $\kappa$ value is high, i.e., the simulated disk has relatively high power in small scale structures.
They deviate significantly (by more than 0.3 dex) from the true profile when $r\gtrsim500''$, close to $R_{\rm HI}$ where the true $\SHI=1~\Msunpcsq$. 
The combined images are better at recovering the true $\SHI$ radial profiles than the VLA images. 
The different types of combined images work almost similarly well out to $r\sim600''$, beyond which the deviations are all large.
It is a bit surprising to see that the standard-combined images, which have bad performance in pixelwise $\SHI$ (and VLA standard images are bad for flux cross-calibration, W24), produce very good radial profiles, possibly because the uncertainties cancel out after the azimuthal averaging. 

\begin{figure*} 
\centering
\includegraphics[width=11cm]{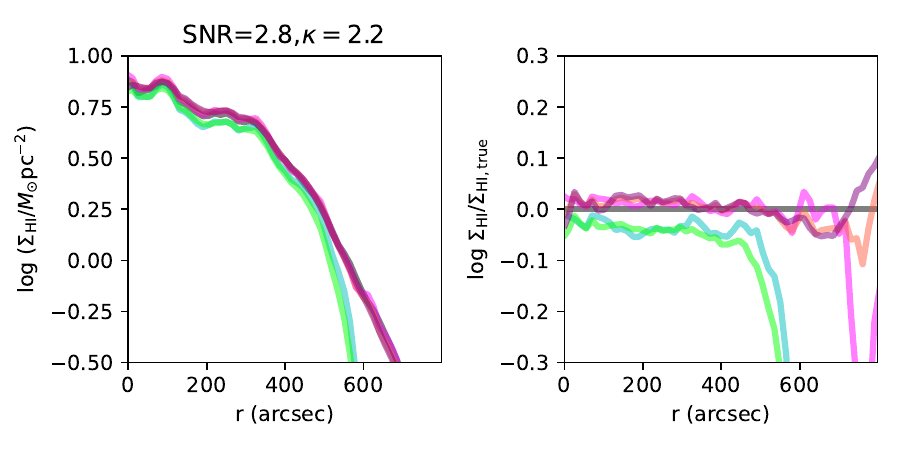}
\includegraphics[width=11cm]{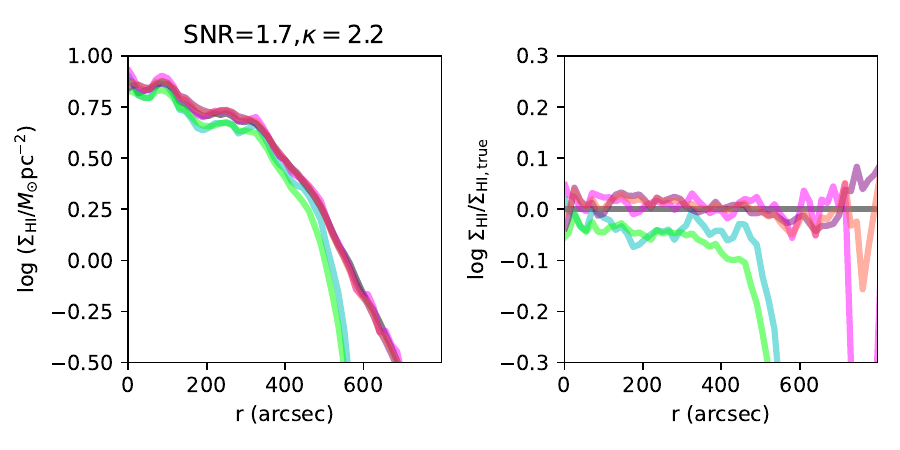}
\includegraphics[width=11cm]{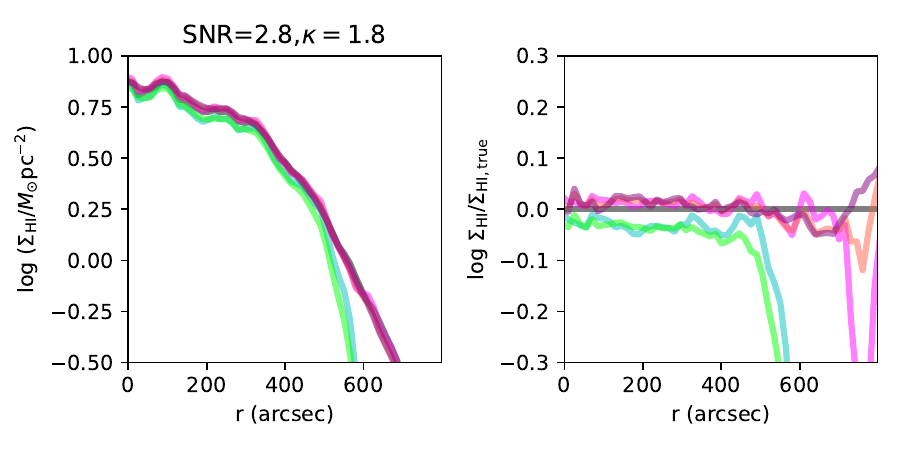}
\includegraphics[width=11cm]{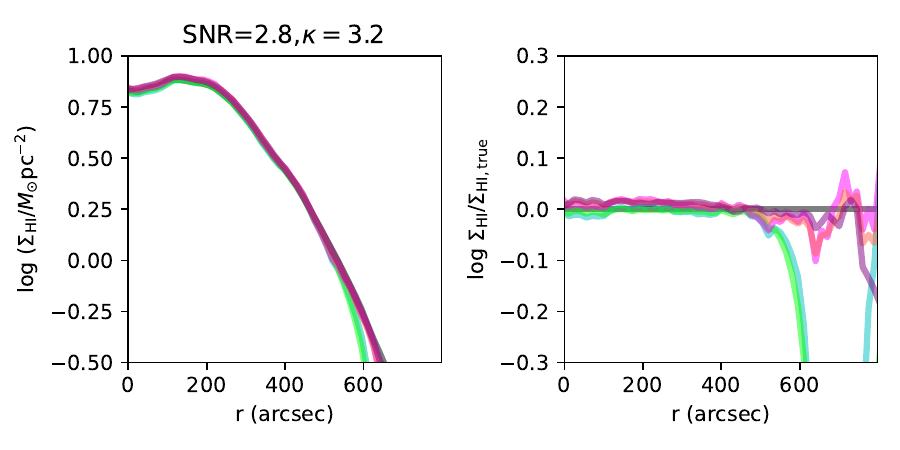}
\caption{ The recovering capabilities of azimuthally averaged $\SHI$ of different types of $\hi$ images. 
The colors and line styles have similar meanings as in Figure~\ref{fig:pixNHI_mock_ref}.  }
\label{fig:profNHI_mock}
\end{figure*}

\subsection{Summary}
\label{sec:summary}
Results from the mock tests suggest that combining with single-dish (FEASTS) image significantly helps recover the true $\SHI$ when the interferometric (THINGS) observation misses fluxes. 
The improvement is robust against a wide range of SNR in the interferometric data, and a wide range of power spectral slopes of $\hi$ disks.
It is true for both azimuthally averaged $\SHI$ and pixelwise $\SHI$. 
While using the standard or residual-rescaled interferometric images for image combination only recovers well the radial profiles of $\SHI$ but not pixelwise measurements of $\SHI$, using convolved models leads to the best recovering of $\SHI$.

\section{Atlas of Combined Images}
\label{app:atals}
We present an atlas of the THINGS and FEASTS combined $\hi$ images. 
The original THINGS and FEASTS images are also displayed for a comparison.

\begin{figure*} 
\centering
\includegraphics[width=5.5cm]{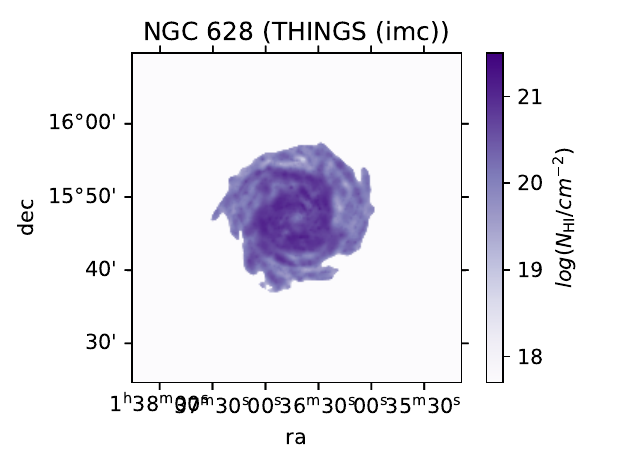}
\includegraphics[width=5.5cm]{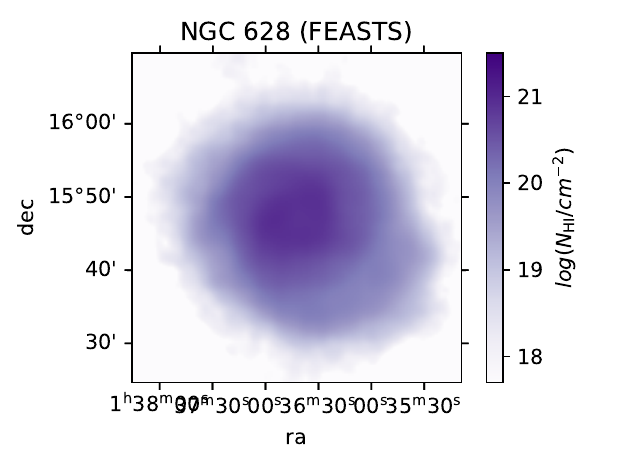}
\includegraphics[width=5.5cm]{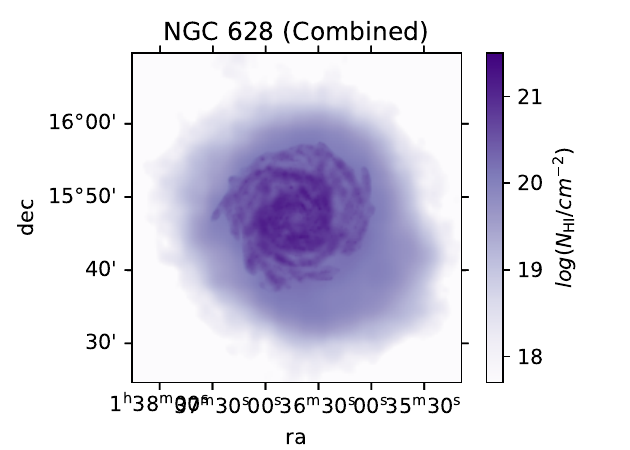}

\includegraphics[width=5.5cm]{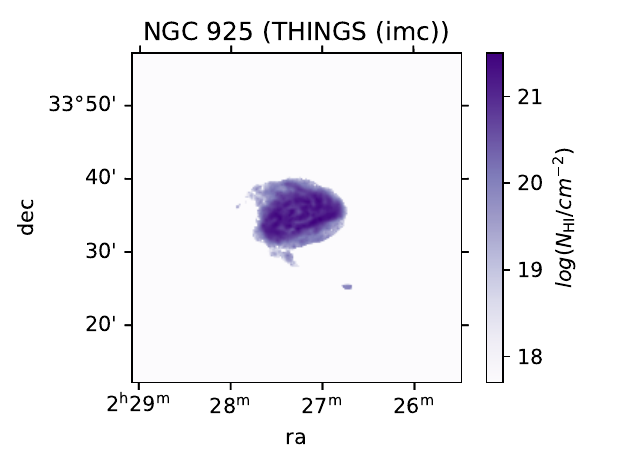}
\includegraphics[width=5.5cm]{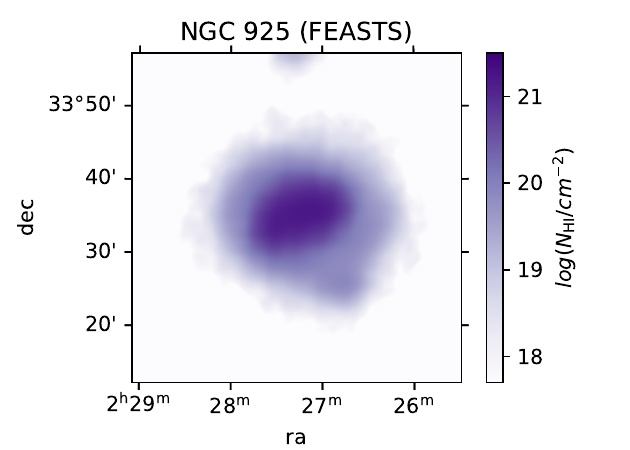}
\includegraphics[width=5.5cm]{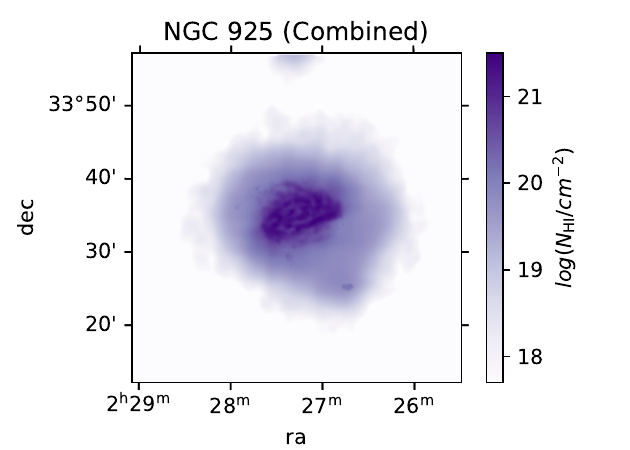}

\includegraphics[width=5.5cm]{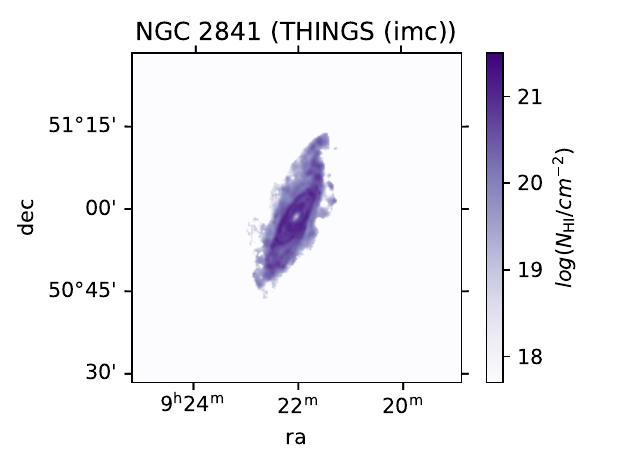}
\includegraphics[width=5.5cm]{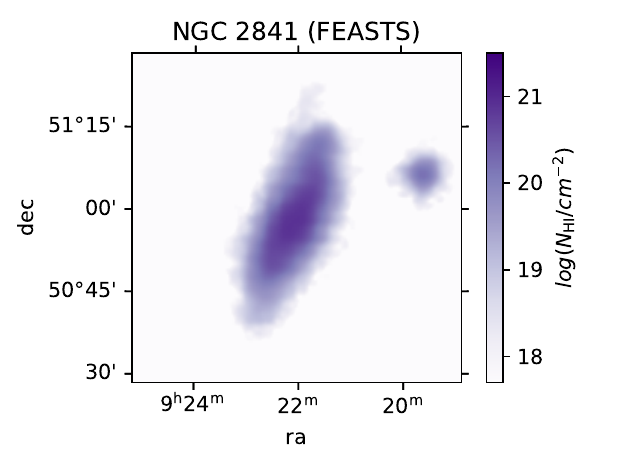}
\includegraphics[width=5.5cm]{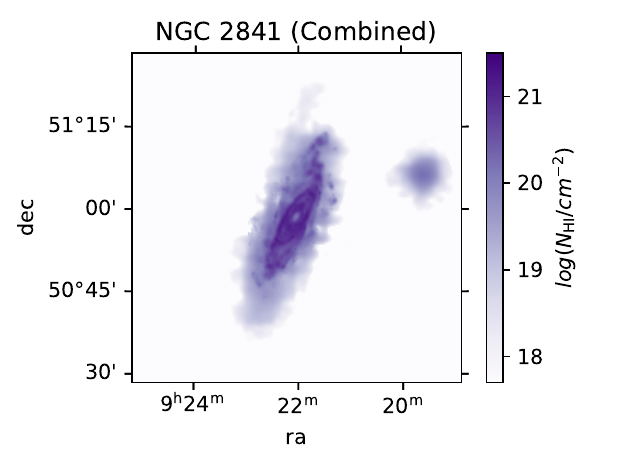}

\includegraphics[width=5.5cm]{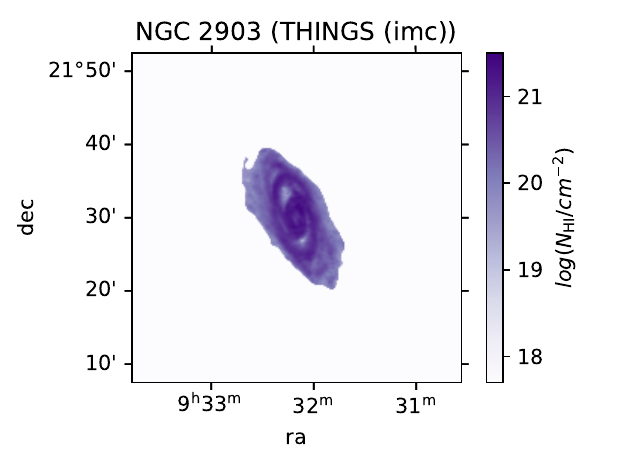}
\includegraphics[width=5.5cm]{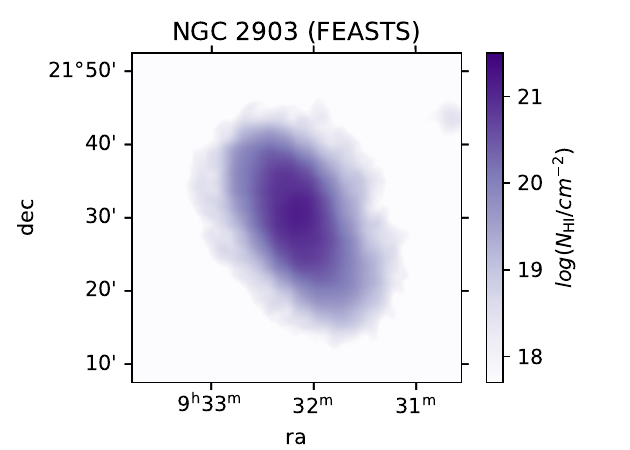}
\includegraphics[width=5.5cm]{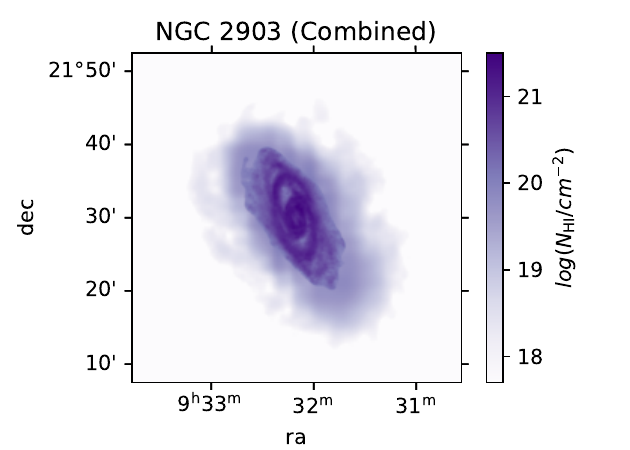}

\includegraphics[width=5.5cm]{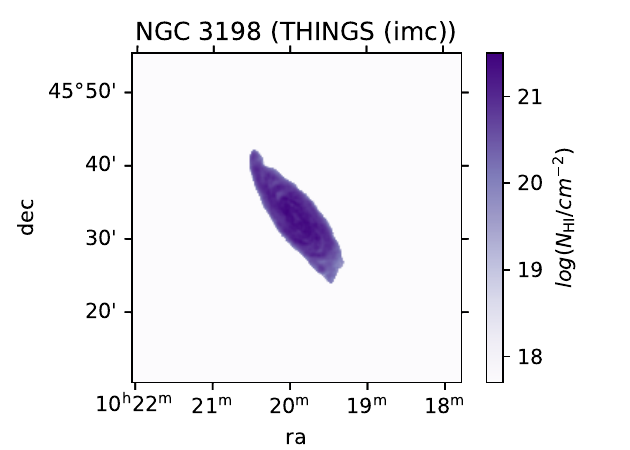}
\includegraphics[width=5.5cm]{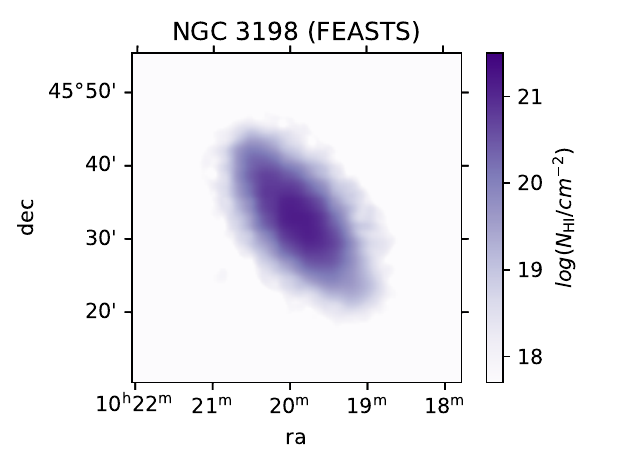}
\includegraphics[width=5.5cm]{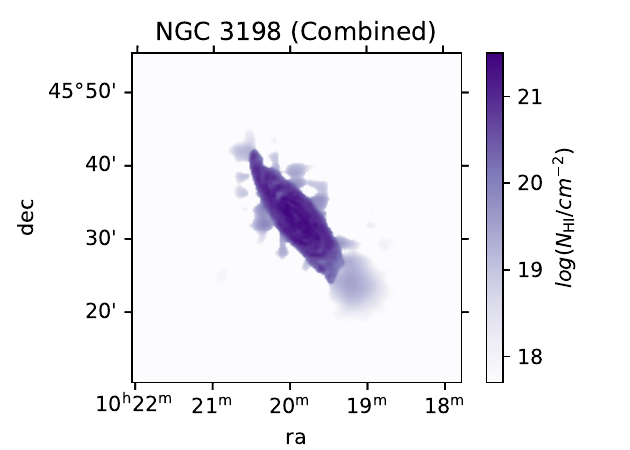}

\caption{Atlas of $\hi$ images with FEASTS observations. 
Each row is for a galaxy, and the three panels are for the original THINGS ($I_\text{W24}$), the FEASTS, and the combined images ($C_\text{W24}$), respectively. 
To be continued.}
\label{fig:atlas}
\end{figure*}

\addtocounter{figure}{-1}

\begin{figure*} 
\centering

\includegraphics[width=5.5cm]{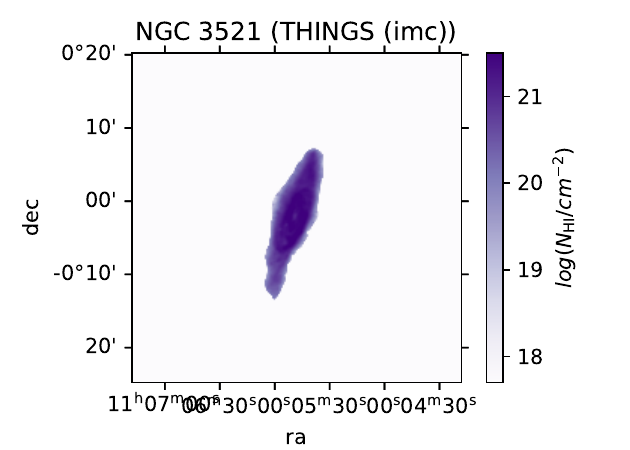}
\includegraphics[width=5.5cm]{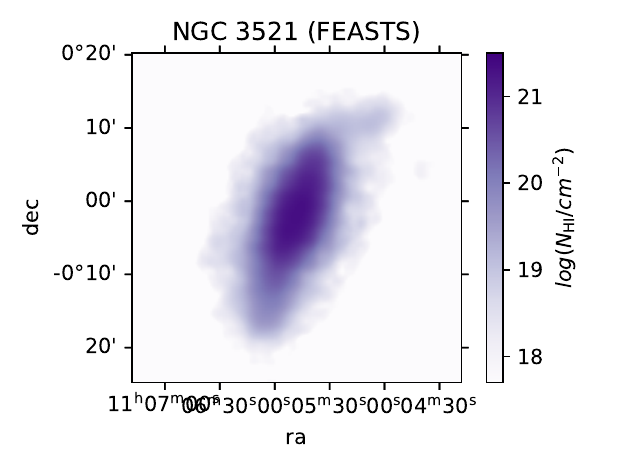}
\includegraphics[width=5.5cm]{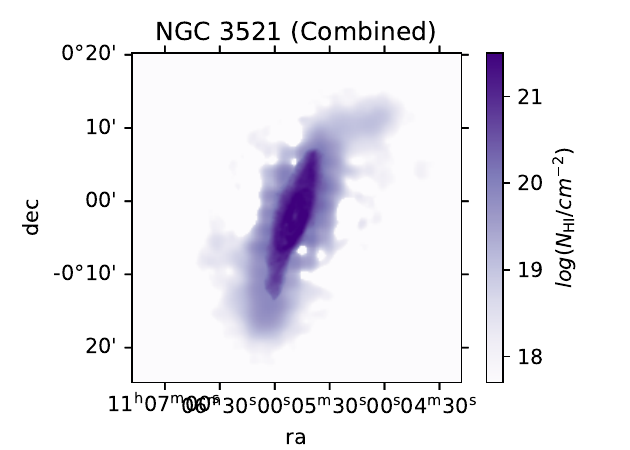}

\includegraphics[width=5.5cm]{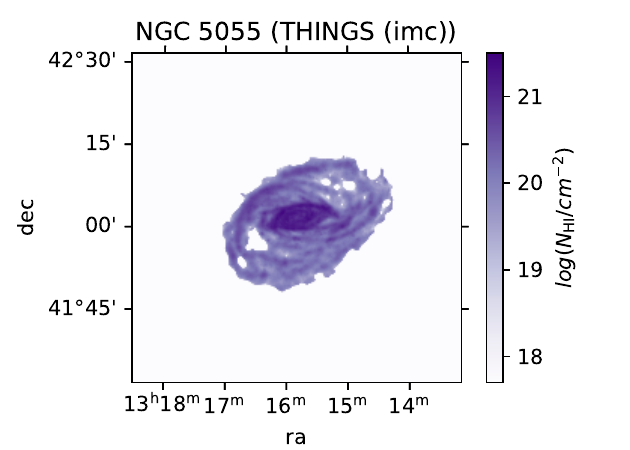}
\includegraphics[width=5.5cm]{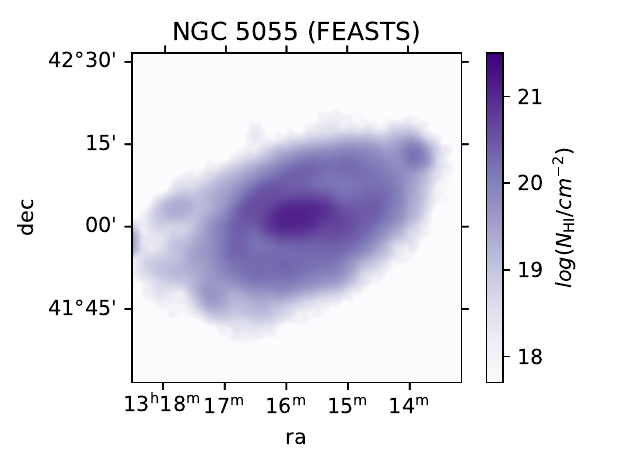}
\includegraphics[width=5.5cm]{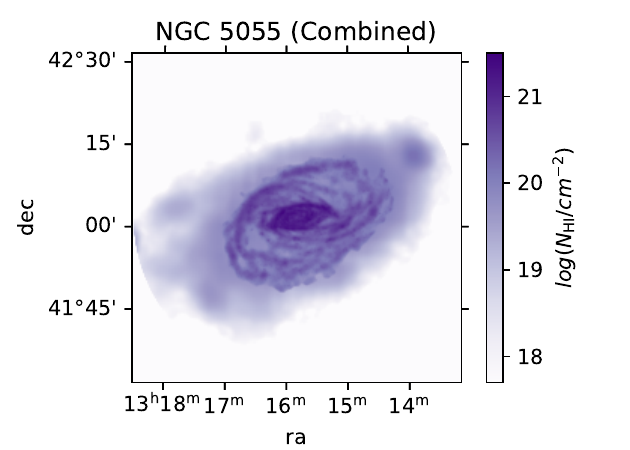}

\includegraphics[width=5.5cm]{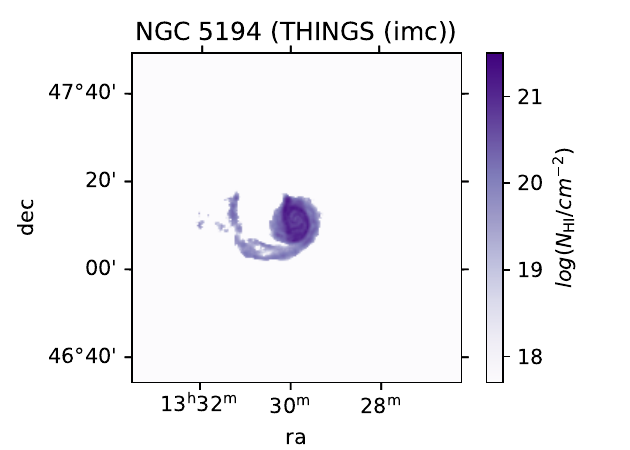}
\includegraphics[width=5.5cm]{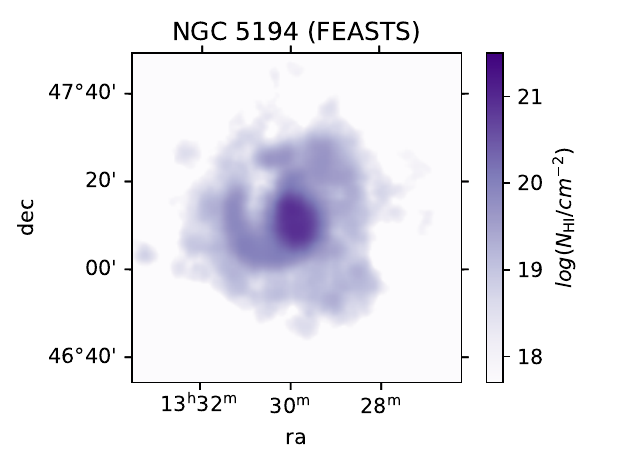}
\includegraphics[width=5.5cm]{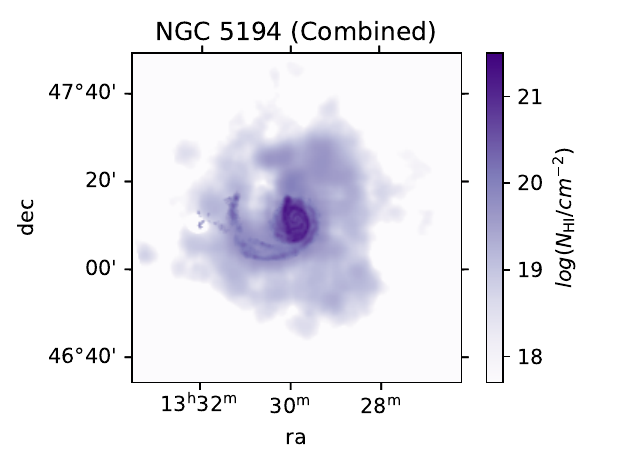}

\includegraphics[width=5.5cm]{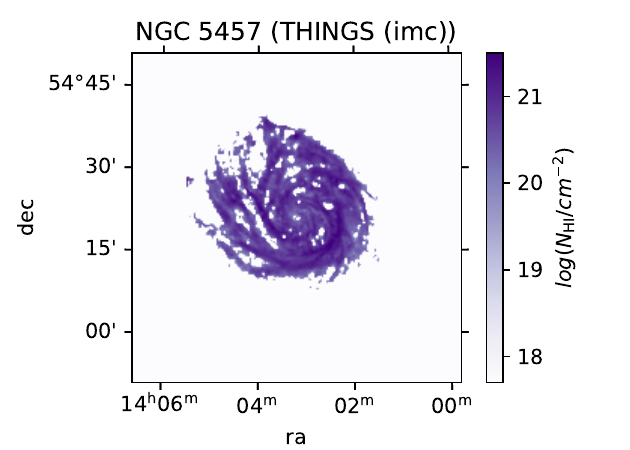}
\includegraphics[width=5.5cm]{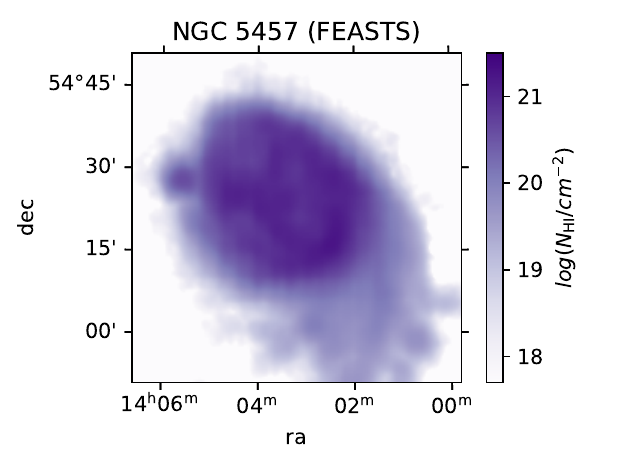}
\includegraphics[width=5.5cm]{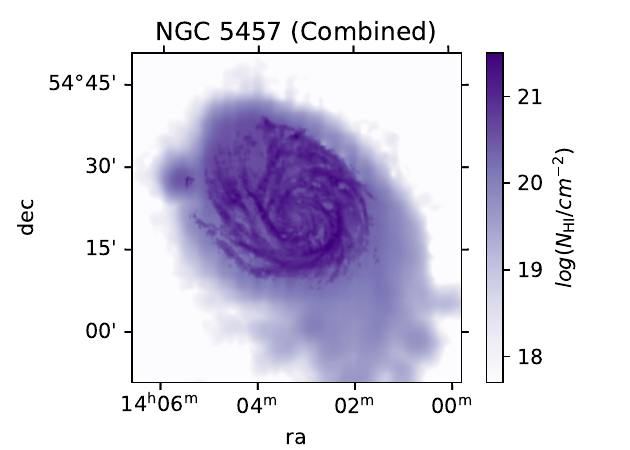}

\includegraphics[width=5.5cm]{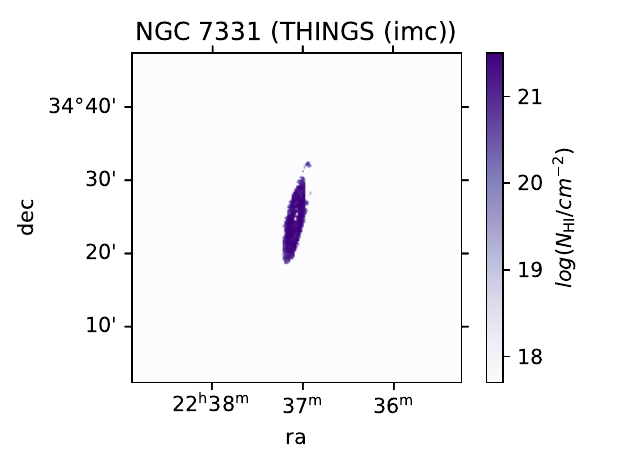}
\includegraphics[width=5.5cm]{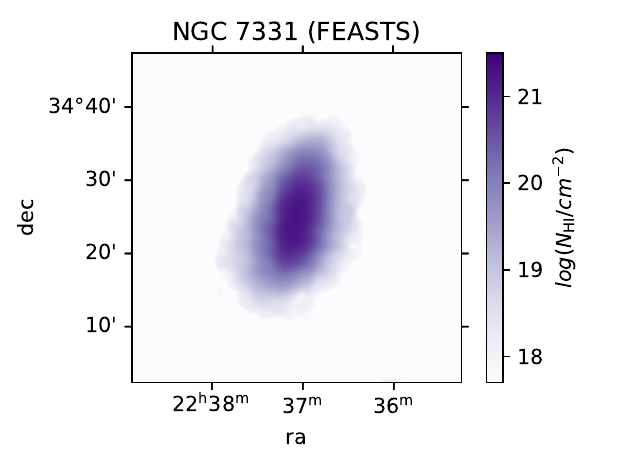}
\includegraphics[width=5.5cm]{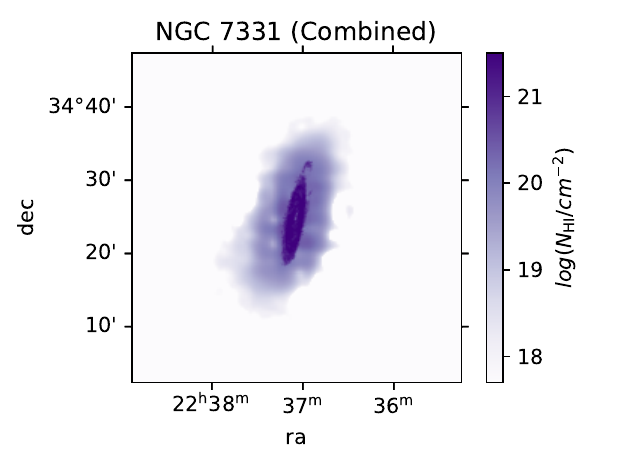}

\caption{Atlas of $\hi$ images with FEASTS observations. Continued.}
\end{figure*}

\bibliography{sfout}{}
\bibliographystyle{apj}

\end{CJK*}
\end{document}